\documentclass[11pt,leqno]{article}
\usepackage{tabularx}
\usepackage{multirow}
\usepackage{eepic,epic}
\usepackage{epsfig}
\usepackage{graphicx}
\usepackage{color}
\usepackage{mathptmx}
\usepackage{amssymb}
\usepackage{amsmath}
\usepackage{pifont}
\usepackage{nicefrac}
\usepackage{graphicx}
\usepackage{caption}
\usepackage[ruled,vlined,linesnumbered]{algorithm2e}
\usepackage{subfigure}

\usepackage[latin1]{inputenc}
\usepackage{longtable}
\usepackage{pifont}
\usepackage{booktabs}
\usepackage{multirow}
\input epsf %

\makeatletter%
\newcommand{\doublespacing}{\let\CS=
\@currsize\renewcommand{\baselinestretch}{1.75}\tiny\CS}
\newcommand{\extradoublespacing}{\let\CS=
\@currsize\renewcommand{\baselinestretch}{1.9}\tiny\CS}
\newcommand{\draftspacing}{\let\CS=
\@currsize\renewcommand{\baselinestretch}{2.0}\tiny\CS}
\newcommand{\hugedraftspacing}{\let\CS=
\@currsize\renewcommand{\baselinestretch}{2.4}\tiny\CS}
\makeatother%

\newcommand{\OMIT}[1]{} %

\newcommand\qedblob{\ding{113}}
\def\literalqed{{\ \nolinebreak\hfill\mbox{\qedblob\quad}}}

\newenvironment{proofs}{\noindent{\bf Proof.}\hspace*{1em}}{\literalqed\bigskip}

\newcommand{\seq}{\subseteq}

\newcommand{\scoreof}[1]{\mathit{score}(#1)}

\newcommand{\scorelev}[1]{\mathit{score}^{#1}}

\newcommand{\maj}[1]{\mathit{maj}(#1)}

\newcommand{\scoresublevel}[3]{\mathit{score}_{#1}^{#2}(#3)}

\newcommand{\weightedmaj}{\mathit{maj}}
\newcommand{\weightedscore}{\mathit{score}}
\newcommand{\weightedscoreofsub}[2]{\mathit{score}_{#1}(#2)}

\newcommand{\p}{\mbox{\rm P}}

\newcommand{\np}{\mbox{\rm NP}}

\newcommand{\condition}{\,|\:}

  \newtheorem{theorem}{Theorem}[section]
  \newtheorem{corollary}[theorem]{Corollary}

  \newtheorem{lemma}[theorem]{Lemma}
  
  \newtheorem{observation}[theorem]{Observation}
  
  \newtheorem{proposition}[theorem]{Proposition}
  \newtheorem{definition}[theorem]{Definition}

\newcommand{\EP}[3]{
\smallskip
\begin{center}
{\small 
\begin{tabularx}{0.9\columnwidth}{ll}
\toprule
\multicolumn{2}{c}{\textsc{#1}} \\
\midrule
{\bf Given:}   & \parbox[t]{0.74\columnwidth}{#2\vspace*{1mm}}  \\
{\bf Question:}& \parbox[t]{0.74\columnwidth}{#3\vspace*{.5mm}} \\ 
\bottomrule
\end{tabularx}
}
\end{center}
\smallskip
}

\usepackage[colorlinks]{hyperref}

\begin{document}

\title{Complexity of Manipulation, Bribery, and Campaign Management in
  Bucklin and Fallback Voting\thanks{This work was supported in part
    by DFG grant RO~1202/15-1, by a DAAD grant for a PPP project in
    the PROCOPE program, by NCN grants 2012/06/M/ST1/00358,
    2011/03/B/ST6/01393, and by the AGH University grant
    11.11.230.015.}
}

\author{Piotr Faliszewski \\
AGH University\\
Krakow, Poland
\and
Yannick Reisch,\ J\"{o}rg Rothe,\ and\ Lena Schend\\
Heinrich-Heine-Universit\"{a}t D\"{u}sseldorf \\
D\"{u}sseldorf, Germany 
}
\date{July 27, 2013}
\maketitle

\hypersetup{
   bookmarksnumbered=true,
   linkcolor=black,
   pdfborderstyle={/S/U/W 1},
   anchorcolor=black,
   citecolor=black,
   filecolor=black,
   urlcolor=black,
   menucolor=black,
   pagecolor=black
}

\begin{abstract}
  A central theme in computational social choice is to study the
  extent to which voting systems computationally resist manipulative
  attacks seeking to influence the outcome of elections, such as
  manipulation (i.e., strategic voting), control, and bribery.
  Bucklin and fallback voting are among the voting systems with the
  broadest resistance (i.e., NP-hardness) to control attacks.
  However, only little is known about their behavior regarding
  manipulation and bribery attacks.  We comprehensively investigate
  the computational resistance of Bucklin and fallback voting for many
  of the common manipulation and bribery scenarios; we also complement
  our discussion by considering several campaign management problems
  for Bucklin and fallback.
\end{abstract}


\section{Introduction}

A central theme in computational social choice (see, e.g., the
bookchapter by Brandt et al.~\cite{bra-con-end:b:comsoc}) is to study
the extent to which voting systems computationally resist manipulative
attacks that seek to influence the outcome of elections, such as
manipulation (i.e., strategic voting), control, and bribery.  In
\emph{manipulation} (introduced by Bartholdi et
al.~\cite{bar-tov-tri:j:manipulating,bar-orl:j:polsci:strategic-voting}
and, more generally, by Conitzer et
al.~\cite{con-san-lan:j:when-hard-to-manipulate}; see, e.g., the
survey by Faliszewski and Procaccia~\cite{fal-pro:j:manipulation}),
voters try to do so by casting insincere votes.  In \emph{control}
(introduced by Bartholdi et al.~\cite{bar-tov-tri:j:control}, see also
Hemaspaandra et al.~\cite{hem-hem-rot:j:destructive-control}), an
election chair tries to influence the election outcome by changing the
structure of the election via adding/deleting/partitioning either
candidates or voters.  In \emph{bribery} (introduced by
Faliszewski~\cite{fal-hem-hem:j:bribery}), an external agent tries to
influence the election outcome by bribing certain voters without
exceeding some given budget.  Since these types of influence are often
possible in principle for many voting systems, it has been studied to
what extent computational hardness can provide some kind of
protection.

Bucklin and fallback
voting~\cite{bra-san:j:preference-approval-voting} are among the
voting systems with the broadest resistance (i.e.,
NP-hardness\footnote{Resistance to manipulative actions is most often
  meant to be NP-hardness in the literature.  Being a worst-case
  measure only, NP-hardness does have its limitations.  There are also
  a number of other approaches that challenge such NP-hardness
  results, surveyed
  in~\cite{rot-sch:j-toappear:survey-typical-case-challenges-manipulation-control};
  for example, there are some experimental results on the control
  complexity of Bucklin and fallback
  voting~\cite{rot-sch:c:fallback-voting-experiments}.})  to control
attacks (see the work of Erd\'{e}lyi et
al.~\cite{erd-rot:c:fallback-voting,erd-pir-rot:c:voter-partition-in-bucklin-and-fallback-voting,erd-fel:c:fallback-voting,erd-fel-pir-rot:t-with-AAMAS-12-pointer:control-in-bucklin-and-fallback-voting}).\footnote{Other
  voting systems whose control complexity has thoroughly been studied
  include plurality, Condorcet, and approval
  voting~\cite{bar-tov-tri:j:control,hem-hem-rot:j:destructive-control},
  Llull and Copeland
  voting~\cite{fal-hem-hem-rot:j:llull-copeland-full-techreport}, a
  variant of approval voting known as
  \mbox{SP-AV}~\cite{erd-now-rot:j:sp-av}, and normalized range
  voting~\cite{men:t-v2:normalized-range-voting-broadly-resists-control}.}
However, only little is known about the behavior of these two voting
systems regarding manipulation and bribery attacks; Schlotter et
al.~\cite{sch-fal-elk:c:campaign-management-under-approval-driven-voting}
have studied them with respect to campaign management, focusing on
shift bribery and support bribery.  We comprehensively investigate the
computational resistance of Bucklin and fallback voting for many of
the common manipulation and bribery scenarios. We also complement the
results of Schlotter et
al.~\cite{sch-fal-elk:c:campaign-management-under-approval-driven-voting}
by studying two other campaign-management problems, namely swap
bribery and extension bribery.

\section{Preliminaries}

\subsection{Bucklin and Fallback Elections}




An \emph{election} is a pair $(C,V)$, where $C = \{c_1, \ldots ,
c_m\}$ is a set of $m$ candidates and $V = (v_1,\ldots,v_n)$ is a list
of votes (or ballots) specifying the $n$ voters' preferences over the
candidates in~$C$.  How these preferences are represented depends on
the voting system used.  We allow voters to be weighted, i.e., a
nonnegative integer weight $w_i$ is associated with each vote~$v_i$.
For example, a vote $v_i$ of a voter with weight $w_i = 3$ is counted
as if three voters with unit weight would have cast the same ballot.
An unweighted election is the special case of a weighted election
where each voter has unit weight.

A voting system is a rule for how to determine the winner(s) of a
given election.  Here we focus on Bucklin and fallback voting only.
Both systems use the notion of \emph{(weighted) majority threshold
  in~$V$}, which is defined by $\weightedmaj(V) = \left\lfloor
\nicefrac{W}{2}\right\rfloor +1$, where $W = \sum_{i=1}^{n} w_i$ is
the total weight of the votes in~$V$.  In \emph{Bucklin voting}, votes
are linear rankings of all candidates, denoted by, e.g., $c_2 > c_3 >
c_1$, which means that this voter (strictly) prefers $c_2$ to $c_3$
and $c_3$ to~$c_1$.  We call the top position in a vote
\emph{level~$1$}, the next position \emph{level~$2$}, and so on.
Starting with the top position and proceeding level by level through
the votes in~$V$, we determine the smallest level $\ell$ such that
some candidate(s) occur(s) in at least $\weightedmaj(V)$ votes up to
this level.\footnote{In \emph{simplified Bucklin voting}, all these
  candidates win.  However, we consider \emph{Bucklin voting} in the
  unsimplified version where winners are determined by a slightly more
  involved procedure.  Note that every Bucklin winner, as defined in
  the main text, also wins in simplified Bucklin voting, but not
  necessarily the other way round.}
A bit more formally, for each candidate $c \in C$, the \emph{Bucklin
  score of $c$ in $(C,V)$}, denoted by $\scoresublevel{(C,V)}{i}{c}$,
is the smallest level $k$ such that $c$ occurs in at least
$\weightedmaj(V)$ votes within the first $k$ levels.  Among the
candidates from $C$ with smallest Bucklin score, say~$\ell$, those
occurring most often up to level $\ell$ are the \emph{Bucklin
  winners}.  If a candidate $c$ becomes a Bucklin winner on level
$\ell$, we sometimes specifically call $c$ a \emph{level $\ell$
  Bucklin winner}.

Fallback voting is a hybrid voting systems designed by Brams and
Sanver~\cite{bra-san:j:preference-approval-voting} to combine Bucklin
with approval voting.  Let us first define approval voting, which was
proposed by Brams and Fishburn~\cite{bra-fis:j:approval-voting} (see
also, e.g.,
\cite{bra-fis:b:approval-voting,bau-erd-hem-hem-rot:b:computational-aspects-of-approval-voting}
for more background).  In \emph{approval voting}, votes in an election
$(C,V)$ are approval vectors from $\{0,1\}^{\|C\|}$ indicating for
each candidate $c \in C$ whether $c$ is approved (``$1$'') by this
voter or not (``$0$'').  Every candidate with the highest approval
score is an \emph{approval winner}.  For each vote $v \in V$, let
$S_v$ denote the \emph{approval strategy of~$v$}, i.e., $S_v \subseteq
C$ contains the candidates approved by~$v$.  In \emph{fallback
  voting}, voters first approve or disapprove of all candidates and
then they provide a linear ranking of all approved candidates.  For
example, some voter might disapprove of $c_1$ and~$c_4$, but approve
of $c_2$ and $c_3$, preferring $c_2$ to~$c_3$; this vote is denoted by
$c_2 > c_3 \,\mid\, \{c_1, c_4\}$.  To determine the winners in
fallback voting, we first try to find the Bucklin winners when they
exist.  If so, all Bucklin winners are \emph{fallback winners}.
However, due to disapprovals it might happen that there is no Bucklin
winner, and in that case all approval winners are \emph{fallback
  winners}.  A bit more formally, given a fallback election $(C,V)$,
let $A(c) = \{v\in V \condition c \in S_v\}$ denote the set of voters
that approve of candidate $c \in C$, let $A^{j}(c)$ denote the set of
voters that approve of candidate $c$ up to the $j$th level, and
define
\[
\weightedscoreofsub{(C,V)}{c} = \sum\limits_{v_i \in A(c)} w_i
\quad\text{ and }\quad
\scoresublevel{(C,V)}{j}{c} = \sum\limits_{v_i \in A^{j}(c)} w_i .
\]
The \emph{fallback score of $c$ in $(C,V)$} is the smallest level $k$
such that $\scoresublevel{(C,V)}{k}{c} \geq \weightedmaj(V)$.  Among
the candidates from $C$ with smallest fallback score, say~$\ell$,
those occurring most often up to level $\ell$ are the
\emph{(level~$\ell$) fallback winners}.  Otherwise (i.e., if no
candidate in $C$ satisfies $\scoresublevel{(C,V)}{k}{c} \geq
\weightedmaj(V)$ for any $k \leq m$), all candidates $c$ with maximum
$\weightedscoreofsub{(C,V)}{c}$ are the \emph{fallback winners}.


It is clear from the definition above that Bucklin elections are special 
fallback elections where all voters approve of all candidates. In other 
scenarios modeling
tampering with election results (e.g., in control scenarios where 
the chair changes the structure of the election
without changing the voters' preferences),
this implies that $\np$-hardness results for control problems in
Bucklin elections can be directly transferred to the same control problems
in the more general fallback elections. 
In manipulation and bribery scenarios, however, such a direct transformation
is not possible because the preferences of certain voters are changed,
and we will show our results separately for both voting systems.

\subsection{Basics from Complexity Theory}

We assume the reader is familiar with the basic notions from
complexity theory such as the complexity classes $\p$ and~$\np$, the
polynomial-time many-one ($\leq_{m}^{p}$) and Turing ($\leq_{T}^{p}$)
reducibility, and with hardness and completeness with respect
to~$\leq_{m}^{p}$.  For more background on complexity theory, see,
e.g., the textbooks~\cite{pap:b:complexity,rot:b:cryptocomplexity}.

\section{Manipulation in Bucklin and Fallback Voting}
\label{sec:manipulation}

\subsection{Definitions and Overview of Results}

Conitzer et al.~\cite{con-san-lan:j:when-hard-to-manipulate}
introduced the following decision problem to model manipulation by a
coalition of weighted voters.  For a given election
system~$\mathcal{E}$, define:

\EP{$\mathcal{E}$-Constructive Coalitional Weighted Manipulation
 ($\mathcal{E}$-CCWM)}
{A set $C$ of candidates, a list $V$ of nonmanipulative votes
  over~$C$ each having a nonnegative integer weight, where $W_V$ is
  the list of these weights, a list $W_S$ of the weights of $k$
  manipulators in $S$ (whose votes over $C$ are still unspecified)
  with $V \cap S = \emptyset$, and a designated candidate $c \in C$.}
{Can the votes in $S$ be set such that $c$ is the unique $\mathcal{E}$
  winner of $(C, V \cup S)$?}
  
The unweighted case $\mathcal{E}$-\textsc{CCUM} is the special case of
$\mathcal{E}$-\textsc{CCWM} where all voters and manipulators have
unit weight.  By changing the question to ``\ldots\ such that $c$ is
not a unique winner in $(C,V\cup S)$?,'' we obtain the destructive
variants, $\mathcal{E}$-\textsc{DCWM} and $\mathcal{E}$-\textsc{DCUM}.
If there is only one manipulator, we denote the corresponding problems
by $\mathcal{E}$-\textsc{CUM}, $\mathcal{E}$-\textsc{CWM},
$\mathcal{E}$-\textsc{DUM}, and $\mathcal{E}$-\textsc{DWM}; these
problems were first studied by Bartholdi et
al.~\cite{bar-tov-tri:j:manipulating,bar-orl:j:polsci:strategic-voting}.

The following proposition follows immediately from the definitions.
  
\begin{proposition}
\label{prop:trivial-reductions-manipulation}
\begin{enumerate}
\item $\mathcal{E}\text{-CUM}\leq_{m}^{p}
\mathcal{E}\text{-CCUM}\leq_{m}^{p}\mathcal{E}\text{-CCWM}$.
\item $\mathcal{E}\text{-DUM}\leq_{m}^{p}
\mathcal{E}\text{-DCUM}\leq_{m}^{p}\mathcal{E}\text{-DCWM}$.
\item $\mathcal{E}\text{-DUM}\leq_{T}^{p}\mathcal{E}\text{-CUM}$, 
$\mathcal{E}\text{-DWM}\leq_{T}^{p}\mathcal{E}\text{-CWM}$.
\item $\mathcal{E}\text{-DCUM}\leq_{T}^{p}\mathcal{E}\text{-CCUM}$.
\item $\mathcal{E}\text{-DCWM}\leq_{T}^{p}\mathcal{E}\text{-CCWM}$.
\end{enumerate}
\end{proposition}


Table~\ref{tab:overview-mani} gives an overview of our results for
manipulation in Bucklin and fallback voting.

\begin{table}
\caption{Overview of results for manipulation in Bucklin and fallback voting}
\label{tab:overview-mani}
 \centering
 \begin{tabular}{@{} l l l l l @{}}
\toprule
				& \multicolumn{2}{c}{Bucklin voting}
                                & \multicolumn{2}{c}{fallback voting} \\
                                & complexity & reference
                                & complexity & reference \\
\cmidrule(r){1-1} \cmidrule(rl){2-3} \cmidrule(l){4-5} 
$\mathcal{E}$-\textsc{CCUM}	& $\p$ & Thm.~\ref{thm:bv-ccum}	
				& $\p$	& Prop.~\ref{prop:fv-unweighted-mani}\\
$\mathcal{E}$-\textsc{DCUM}	& $\p$ & Cor.~\ref{cor:bv-unweighted-mani}		
				& $\p$	& Prop.~\ref{prop:fv-unweighted-mani}\\
$\mathcal{E}$-\textsc{CCWM}	& $\np$-complete & Thm.~\ref{thm:bv-ccwm-nphard}	
				& $\p$	& Prop.~\ref{prop:fv-weighted-mani}\\
$\mathcal{E}$-\textsc{DCWM}	& $\p$ & Thm.~\ref{thm:bv-dcwm-p} 
				& $\p$	& Prop.~\ref{prop:fv-weighted-mani}\\

\bottomrule	 
\end{tabular}
\end{table}

\subsection{Results for Unweighted Manipulation}
\label{sec:unweighted-mani}

 
In fallback elections, manipulators that try to make a certain candidate 
the winner by changing their votes can follow a simple strategy: 
They can limit their approval strategy to only this candidate and thus 
preclude all other candidates from gaining points from their votes. 
It is easy to see that if this attempt is not successful, no other 
way of constructing the manipulators' votes can make their designated 
candidate win. 
This means that fallback-\textsc{CCUM} is in $\p$, which implies $\p$-membership 
for fallback-\textsc{CUM}, fallback-\textsc{DCUM}, and fallback-\textsc{DUM} as well 
(with Proposition~\ref{prop:trivial-reductions-manipulation}). 
We state this observation in the following proposition.

\begin{proposition}
\label{prop:fv-unweighted-mani}
Fallback-\textsc{CCUM}, fallback-\textsc{CUM}, fallback-\textsc{DCUM},
and fallback-\textsc{DUM} are in~$\p$.
\end{proposition}

In Bucklin elections, however, the argumentation is more involved, since 
the manipulators do not have the possibility to preclude any candidate 
from gaining points from their votes. So the manipulators' votes have to 
be carefully constructed to ensure that no other candidate than the designated 
candidate gains too much points on the relevant levels. 

Nevertheless, we can show that 
Bucklin-\textsc{CCUM} is in $\p$ by adapting an algorithm
for \emph{simplified-Bucklin}-\textsc{CCUM} that is due to Xia
et al.~\cite{xia-zuc-pro-con-ros:c:unweighted-coalitional-manipulation},
see Algorithm~\ref{alg:bv-ccum}.

  %
\begin{algorithm} 
\small
\DontPrintSemicolon
\SetKwInOut{Input}{input}\SetKwInOut{Output}{output}
\Input{$C$ set of candidates\\
~$V$ list of voters\\  
~$k$ number of manipulators \\
~$p$ designated candidate}
\Output{``YES'' if $(C,V,k,p)\in\text{Bucklin-\textsc{CCUM}}$ \\
~``NO'' if $(C,V,k,p)\notin\text{Bucklin-CCUM}$}
\BlankLine
\uIf{$k>\|V\|$}{return ``YES'';}
let $rem$, $rem_2$, $num$, $num_2$ be arrays of length m;\;
$\mathit{maj}=\lfloor \frac{\|V\|+k}{2}\rfloor+1$;\;
$r_{\min}=\min\{i \condition 
\scoresublevel{(C,V)}{i}{p}+k\geq \mathit{maj}\}$;\;
$S=\text{list of manipulators};$\;
\ForEach{$c\in C-\{p\}$}
{\uIf{$\min\{i \condition 
\scoresublevel{(C,V)}{i}{c}\geq \mathit{maj}\}<r_{\min}$ \textup{OR} 
$\scoresublevel{(C,V)}{r_{\min}}{c}\geq \scoresublevel{(C,V)}{r_{\min}}{p}+k$}
{return ``NO'';}
$rem[c]=\scoresublevel{(C,V)}{r_{\min}}{p}+k-\scoresublevel{(C,V)}{r_{\min}}{c}-1$;\;
$rem_2[c]=\mathit{maj}-\scoresublevel{(C,V)}{r_{\min}-1}{c}-1$;\;
$num[c]=\min\{rem_2[c], rem[c], k\};$\;
$num_2[c]=\min\{rem[c], k\};$\;
}
\uIf{$\sum_{c\in C-\{p\}} \min\{rem_2[c], 
rem[c], k\} < (r_{\min}-2) k$ \textup{OR} $\sum_{c\in C-\{p\}}
\min\{rem[c], k\} < (r_{\min}-1) k$}
{return ``NO'';}
let $tmp_1,\ldots,tmp_k$ represent the manipulators' votes (empty at the beginning);\;
put $p$ on the first position in all the votes of the manipulators;\;
$i=1$;\;
$j=2$;\;
\ForEach{$c\in C-\{p\}$}
{
\While{$num[c]>0$ \textup{AND} $j\leq r_{\min}-1$}{$tmp_i=tmp_i+c$;\;
$num[c]--;$\; $num_2[c]--;$\; $i++;$\; \uIf{$i==k+1$}{$i=1$;\; $j++$;}}
}
\ForEach{$c\in C-\{p\}$}
{
\While{$num_2[c]>0$ \textup{AND} $j==r_{\min}$}{$tmp_i=tmp_i+c$;\;
fill the remaining positions of $tmp_i$ arbitrarily;\; 
$S=S+tmp_i;$\; $num_2[c]--;$\; $i++;$\; \uIf{$i==k+1$}{$j++$;}}
}
return ``YES'';
\caption{Algorithm for Bucklin-\textsc{CCUM}}
\label{alg:bv-ccum}
\end{algorithm}
Before we prove that the presented algorithm is correct and in~$\p$,
we show the following useful lemma.

\begin{lemma}
\label{lem:ccum}
Considering the notation $C$, $V$, $k$, $p$, $rem$, $rem_{2}$,
$num$, $num_{2}$, $r_{\min}$, $S$, and $\mathit{maj}$ as in 
Algorithm~\ref{alg:bv-ccum}, it holds that:
\begin{enumerate}
\item If $k>\|V\|$ then $(C,V,k,p)\in$ Bucklin-\textsc{CCUM}.
\item If there is a candidate $c\in C-\{p\}$ with
\begin{enumerate}
 \item $\min\{i \condition\scoresublevel{(C,V)}{i}{c}\geq \mathit{maj}\}<r_{\min}$ or
 \item $\scoresublevel{(C,V)}{r_{\min}}{c}\geq 
\scoresublevel{(C,V)}{r_{\min}}{p}+k$, 
\end{enumerate}

then $(C,V,k,p)\notin$ Bucklin-\textsc{CCUM}. 
\item $(C,V,k,p)\notin$ Bucklin-\textsc{CCUM} if and only if  

\begin{enumerate}
 \item $\sum_{c\in C-\{p\}}\min\{rem[c],rem_{2}[c],k\}<(r_{\min}-2) k$ or
 \item   $\sum_{c\in C-\{p\}}\min\{rem[c],k\}<(r_{\min}-1) k$.
\end{enumerate}
\end{enumerate}

\end{lemma}

\begin{proofs}
Note that 
$r_{\min}$ denotes the smallest level on which candidate $p$ 
reaches the majority threshold $\mathit{maj}$ in the manipulated election 
assuming that all manipulators position $p$ on the first place. 
So $r_{\min}$ is the smallest level on which $p$ can win. 
This implies that $\scoresublevel{(C,V)}{r_{\min}}{p}+k$ is the 
number of points $p$ has to win the election with. 
Now we can show the three claims. 

\begin{enumerate}
 \item If the number of manipulators is bigger than the number of 
 truthful voters, a successful manipulation is always possible. 
 The manipulators simply position $p$ on the 
 first place in their vote and $p$ reaches the majority threshold 
 already on the first level. So $(C,V,k,p)\in$ Bucklin-\textsc{CCUM} 
 trivially holds.
 
 \item Let $c\in C-\{p\}$ be an arbitrary candidate. 
 
 \begin{enumerate}
  \item It holds that $\min\{i \condition
    \scoresublevel{(C,V)}{i}{c}\geq \mathit{maj}\}<r_{\min}$: That
    means that we have a candidate $c$ that reaches $\mathit{maj}$
    votes on an earlier level than $p$ and $c$ does so even without
    the manipulators' votes.  Thus $(C,V,k,p)\notin$
    Bucklin-\textsc{CCUM}.
  \item It holds that  $\scoresublevel{(C,V)}{r_{\min}}{c}\geq 
\scoresublevel{(C,V)}{r_{\min}}{p}+k$: This means that $c$ gets 
  at least as many points from the truthful voters on the exact level $p$ would 
  have to win the manipulated election as $p$ gains in the election where 
  the manipulators' votes have already been added. That means that 
  $p$ cannot be made the unique winner of the manipulated election and thus 
   $(C,V,k,p)\notin$ Bucklin-\textsc{CCUM} holds. 
 \end{enumerate}
 
 \item 
The array $rem$ indicates for every candidate $c$ how many further
points $c$ can gain without exceeding
$\scoresublevel{(C,V)}{r_{\min}}{p}$ on level~$r_{\min}$.  The array
$rem_2$, on the other hand, indicates for every candidate $c$ how many
further points $c$ may gain without exceeding $\mathit{maj}$ on the
levels~$1$ to~$(r_{\min}-1)$.  For all candidates, $rem$ and $rem_2$
contain positive numbers.  Since every candidate can gain $k$ points
from the manipulators' votes, $num[c]=\min\{rem[c],rem_{2}[c],k\}$ is
the number of manipulators that may have candidate $c$ in the first
$(r_{\min}-1)$ positions of their votes without preventing $p$ from
winning.  Analogously, $num_{2}[c]=\min\{rem[c],k\}$ is the number of
manipulators that can place $c$ among their top $r_{\min}$ positions
without preventing $p$ from winning.  We have that $num_{2}[c]\geq
num[c]$ for all $c\in C-\{p\}$.  We now show the equivalence.

From right to left: 
\begin{enumerate}
 \item Suppose that $\sum_{c\in
   C-\{p\}}\min\{rem[c],rem_{2}[c],k\}<(r_{\min}-2) k$.  In this case,
   it is not possible to fill the remaining $(r_{\min}-2)k$ positions
   (positions~$2$ to~$(r_{\min}-1)$) in the manipulators' votes
   without having for at least one candidate $d\in C-\{p\}$ that
   either
 \begin{eqnarray*}
rem_{2}[d]-\scoresublevel{(C,S)}{r_{\min}-1}{d}<0 &  & \text{ or }\\
rem[d]-\scoresublevel{(C,S)}{r_{\min}}{d}<0
\end{eqnarray*}
holds.  That is equivalent to either
\begin{eqnarray*}
\mathit{maj}-\scoresublevel{(C,V)}{r_{\min}-1}{d}-1-\scoresublevel{(C,S)}{r_{\min}-1}{d}
<0 &  & \text{ or }\\
\scoresublevel{(C,V)}{r_{\min}}{p}+k-\scoresublevel{(C,V)}{r_{\min}}{d}
-1-\scoresublevel{(C,S)}{r_{\min}}{d}<0,
\end{eqnarray*}
which in turn is equivalent to either
\begin{eqnarray*}
\scoresublevel{(C,V\cup S)}{r_{\min}-1}{d}=\scoresublevel{(C,V)}{r_{\min}-1}{d}
+\scoresublevel{(C,S)}{r_{\min}-1}{d}
>\mathit{\mathit{maj}}-1 &  & \text{ or }\\
\scoresublevel{(C,V\cup S)}{r_{\min}}{d}
=\scoresublevel{(C,V)}{r_{\min}}{d}
+\scoresublevel{(C,S)}{r_{\min}}{d}
>\scoresublevel{(C,V)}{r_{\min}}{p} +k-1.
\end{eqnarray*}

So we have that either $d$ is a Bucklin winner in the manipulated
election on a smaller level than~$r_{\min}$, or it holds that on level
$r_{\min}$ candidate $d$ might have at least as many points as $p$.
Thus $(C,V,k,p)\notin$ Bucklin-\textsc{CCUM}.

\item Suppose that $\sum_{c\in C-\{p\}}\min\{rem[c],k\}<(r_{\min}-1)
  k$.  In this case, it is not possible to fill the remaining
  $(r_{\min}-1)k$ positions (positions~$2$ to~$r_{\min}$) in the
  manipulators' votes without having for at least one candidate $d\in
  C-\{p\}$:
\begin{eqnarray*}
& & rem[d]-
\scoresublevel{(C,S)}{r_{\min}}{d}<0 \\
& \Leftrightarrow &
\scoresublevel{(C,V)}{r_{\min}}{p}
+k-
\scoresublevel{(C,V)}{r_{\min}}{d}
-1-
\scoresublevel{(C, S)}{r_{\min}}{d}
<0 \\
& \Leftrightarrow &
\scoresublevel{(C,V\cup S)}{r_{\min}}{d}
=
\scoresublevel{(C,V)}{r_{\min}}{d}
+
\scoresublevel{(C,S)}{r_{\min}}{d}
>
\scoresublevel{(C,V)}{r_{\min}}{p}
+k-1.
\end{eqnarray*}
So we have that $d$ can have at least as many points as $p$ on
level~$r_{\min}$.  So $(C,V,k,p)\notin$ Bucklin-\textsc{CCUM}.
\end{enumerate}

From left to right: We show the contrapositive. Assume
that both
\begin{enumerate}
 \item $\sum_{c\in C-\{p\}}\min\{rem[c],rem_{2}[c],k\} \geq (r_{\min}-2) k$ and
 \item   $\sum_{c\in C-\{p\}}\min\{rem[c],k\} \geq (r_{\min}-1) k$
\end{enumerate}
 hold. Then we can fill positions~$2$ to~$r_{\min}$ of the manipulators' 
 votes in a way that for all candidates $e\in C-\{p\}$ the following holds:
 \begin{eqnarray*}
rem_{2}[e]-
\scoresublevel{(C,S)}{r_{\min}-1}{e}
\geq0 &  & \text{and}\\
rem[e]-
\scoresublevel{(C,S)}{r_{\min}}{e}
\geq0,  &  & \text{}
\end{eqnarray*}
which is equivalent to 
\begin{eqnarray*}
\scoresublevel{(C,V\cup S)}{r_{\min}-1}{e}
=
\scoresublevel{(C,V)}{r_{\min}-1}{e}
+
\scoresublevel{(C, S)}{r_{\min}-1}{e}
\leq \mathit{\mathit{maj}}-1 &  & \text{and}\\
\scoresublevel{(C,V\cup S)}{r_{\min}}{e}
=
\scoresublevel{(C,V)}{r_{\min}}{e}
+
\scoresublevel{(C,S)}{r_{\min}}{e}
\leq 
\scoresublevel{(C,V)}{r_{\min}}{p}
+k-1.
\end{eqnarray*}

So we have that $(C,V,k,p)\in$ Bucklin-\textsc{CCUM}. 

\end{enumerate}
This completes the proof.~\end{proofs}

Now we are ready to show that Algorithm~\ref{alg:bv-ccum} is in $\p$
and correctly solves Bucklin-\textsc{CCUM}.

\begin{theorem}
\label{thm:bv-ccum}
Algorithm~\ref{alg:bv-ccum} has a runtime of $\mathcal{O}(m^{2}+nm)$
and decides Bucklin-\textsc{CCUM}, and thus this problem is in~$\p$.
\end{theorem}

\begin{proofs}
  It follows immediately from Lemma~\ref{lem:ccum} that
  Algorithm~\ref{alg:bv-ccum} is correct.
%
  It is also clear that it always terminates.  To compute the needed
  scores $\scoresublevel{(C,V)}{i}{c}$ for all candidates $c$ and
  every level~$i$, $\mathcal{O}(m^2+nm)$ steps are needed.  The
  for-loop in line~$7$ needs $\mathcal{O}(m)$ steps, while the other
  two for-loops in lines~$20$ and~$30$ need $\mathcal{O}(km)$ steps.
  Since the loops are only run through when $k\leq n$, we have a
  runtime of $\mathcal{O}(nm)$ for the loops, which implies that the
  algorithm has a runtime of $\mathcal{O}(m^2+nm)$ in
  total.~\end{proofs}

With Theorem~\ref{thm:bv-ccum} and
Proposition~\ref{prop:trivial-reductions-manipulation} we have the
following corollary.

\begin{corollary}
Bucklin-\textsc{CUM}, Bucklin-\textsc{DUM}, Bucklin-\textsc{CCUM}, and
Bucklin-\textsc{DCUM} are in $\p$.
\label{cor:bv-unweighted-mani}
\end{corollary}

\subsection{Results for Weighted Manipulation}
\label{sec:weighted-mani}

In this section we analyze the complexity of weighted manipulation in
Bucklin 
and fallback 
voting. 

With the same argumentation as that given at the beginning of
Section~\ref{sec:unweighted-mani}, it is easy to see that in fallback
elections the weighted manipulation problems can be decided
efficiently, namely in deterministic polynomial time: In the
constructive, coalitional, weighted case, all the manipulators need to
do is to approve of the designated candidate---if this attempt does
not lead to the desired result, no other way of changing the
manipulators' votes will.  Again, with
Proposition~\ref{prop:trivial-reductions-manipulation}, the result for
this case directly transfers to the constructive, weighted case with
only one manipulator, and from these two constructive cases to the
corresponding destructive cases.  We state this observation in the
following proposition.

\begin{proposition}
\label{prop:fv-weighted-mani}
Fallback-\textsc{CCWM}, fallback-\textsc{CWM}, fallback-\textsc{DCWM},
and fallback-\textsc{DWM} are in~$\p$.
\end{proposition}

In weighted Bucklin elections, on the other hand, a coalition of manipulators 
trying to make a certain candidate win 
is faced with a harder challenge, as the following result shows.


\begin{theorem}
\label{thm:bv-ccwm-nphard}
Bucklin-\textsc{CCWM} is $\np$-complete.
\end{theorem}


\begin{proofs}
It is easy to see that Bucklin-\textsc{CCWM} is in~$\np$.
We show $\np$-hardness of this problem by a reduction from the
problem \textsc{Partition}: Given a set $A=\{1,\ldots , k\}$ and a
sequence $(a_1, \ldots, a_k)$ of nonnegative integers with
$\sum_{i=1}^{k}a_i = 2K$ for a positive integer~$K$, is there a
set $A' \seq A$ such that $\sum_{i\in A'}a_i =
\sum_{i\not\in A'}a_i = K$?  \textsc{Partition} is well-known to
be $\np$-complete (see, e.g., \cite{gar-joh:b:int}).

Let an instance of \textsc{Partition} be given by
$A=\{1,2,\ldots,k\}$ and $(a_{1},\ldots,a_{k})$ with
$\sum_{i=1}^{k}a_{i}=2K$.
Without loss of generality, we may assume that $a_{i}\geq2$ for each
$i\in A$.  We construct the following instance of
Bucklin-\textsc{CCWM}.  The candidate set is
$C=\{b,c,d,p\}$ and $p$ is the designated candidate.  In $V$ we have
three voters of the following form with a total weight of $6K-2$:
\begin{enumerate}
\item $c>p>d>b$ with weight $2K$,
\item $c>d>p>b$ with weight $K-1$,
\item $b>d>p>c$ with weight $3K-1$,
\end{enumerate}
so the majority threshold in $(C,V)$ is reached with
$\lfloor\nicefrac{(6K-2)}{2}\rfloor+1=3K$.  For the first two levels,
the scores of the candidates are given in
Table~\ref{tab:bv-np-hard-CCWM-scores-(C,V)}, and the unique level~$2$
Bucklin winner in $(C,V)$ is~$d$.
Furthermore, there are $k$ manipulators in $S$ with weights
$a_{1},a_{2},\ldots,a_{k}$, which are given in our
Bucklin-\textsc{CCWM} instance.

\begin{table}[h!t]
\caption{Level~$i$ scores in $(C,V)$ for $i \in \{1,2\}$ and
 the candidates in~$C$}
\label{tab:bv-np-hard-CCWM-scores-(C,V)}
\centering
\begin{tabular}{@{} l c c c c @{}}
\toprule
		& $b$ 		& $c$	 & $d$	& $p$	\\
\cmidrule(r){1-1} \cmidrule(rl){2-2} \cmidrule(rl){3-3} \cmidrule(l){4-4} \cmidrule(l){5-5}
$\scorelev{1}$	& $3K-1$	& $3K-1$ & $0$	  & $0$   \\
$\scorelev{2}$ 	& $3K-1$	& $3K-1$ & $4K-2$ & $2K$  \\
\bottomrule	 
\end{tabular}
\captionsetup{font=small}
\end{table}

We claim that $(A,(a_1,a_2,\ldots,a_k))\in$ \textsc{Partition} if and
only if $p$ can be made the unique Bucklin winner in $(C,V\cup S)$.

From left to right: Assume that there is a subset $A'\seq A$ with
$\sum_{i\in A'}a_{i}=K$.  The majority threshold in $(C,V\cup S)$ is
$\lfloor\nicefrac{(6K-2+2K)}{2}\rfloor+1=4K$.  Let the votes of the
manipulators be of the following form
\begin{itemize}
\item $p>c>b>d$ for all manipulators with weight $a_{i}$ for $i\in A'$,
  and
\item $p>b>c>d$ for the remaining manipulators.
\end{itemize}

For the first two levels, the scores of the candidates in $(C,V\cup
S)$ are given in Table~\ref{tab:bv-np-hard-CCWM-scores-(C,VS)}.  It
follows that $p$ is the unique level~$2$ Bucklin winner in $(C,V\cup S)$.

\begin{table} 
\caption{Level~$i$ scores in $(C,V\cup S)$ for $i \in \{1,2\}$ and
 the candidates in~$C$}
\label{tab:bv-np-hard-CCWM-scores-(C,VS)}
\centering
\begin{tabular}{@{} l c c c c @{}}
\toprule
		& $b$ 		& $c$	 & $d$	& $p$	\\
\cmidrule(r){1-1} \cmidrule(rl){2-2} \cmidrule(rl){3-3} \cmidrule(l){4-4} \cmidrule(l){5-5}
$\scorelev{1}$	& $3K-1$	& $3K-1$ & $0$	& $2K$   	\\
$\scorelev{2}$ 	& $4K-1$	& $4K-1$ & $4K-2$	& $4K$		\\
\bottomrule	 
\end{tabular}
\captionsetup{font=small}
\end{table}

From right to left: Assume that there are votes for the manipulators
in $S$ that make $p$ the unique winner of $(C,V\cup S)$. Without loss
of generality, assume that $p$ is on the first position in all votes
in~$S$.  Note that $p$ cannot win the manipulated election on the
first level, so $p$ has to be the unique level~$2$ winner with
$\scoresublevel{(C,V\cup S)}{2}{p}=4K$.  This implies that
$\scoresublevel{(C,V\cup S)}{2}{e}<4K$ has to hold for all $e\in
\{b,c,d\}$.  Since $a_{i}\geq2$, candidate $d$ cannot be on the second
position in any manipulator's vote. Thus, the manipulators' votes can
be only of the form $(p>c>b>d)$, $(p>c>d>b)$, $(p>b>c>d)$, or
$(p>b>d>c)$.  The candidates $b$ and $c$ have already $3K-1$ points on
the second level in $(C,V)$, so they each cannot gain more than $K$
points on the second level from the votes in~$S$.  Since all votes in
$S$ have either $b$ or $c$ on the second position, the weights of the
manipulators have to be of the form that a subset $A'\seq A$ can be
found such that those manipulators with weight~$a_i$, $i\in A'$, have
a total weight of $K$ and put one of $b$ and $c$ (say $b$) on the
second position, and the remaining manipulators (those with weight
$a_i$ for $i\not\in A'$) put the other candidate, $c$, on the second
position and have a total weight of $K$ as well.  Thus,
$(A,(a_1,a_2,\ldots,a_k))$ is a yes-instance of
\textsc{Partition}.~\end{proofs}

%

We now turn to the desctructive variant of coalitional weighted
manipulation and give a deterministic polynomial-time algorithm for
this problem in Bucklin voting.

\begin{algorithm}   
\DontPrintSemicolon
\SetKwInOut{Input}{input}\SetKwInOut{Output}{output}
\Input{$C$ set of candidates\\
~$V$ list of voters  \\  
~$W_V$ weights of the voters\\
~$W_S$ weights of the manipulators \\ 
~$p$ designated candidate}

\Output{``YES'' if $(C,V,W_V,W_S,p)\in$ Bucklin-\textsc{DCWM}\\
~``NO'' if  $(C,V,W_V, W_S,p)\notin$ Bucklin-\textsc{DCWM}}
\BlankLine
\uIf{$\sum_{w\in W_S} w>\sum_{w\in W_V} w$}{return ``YES'';\;}
\ForEach{$c\in C-\{p\}$}
{put $p$ on the last position in the manipulators' votes;\;
 put $c$ on the first position in the manipulators' votes;\;
fill the remaining positions in the manipulators' votes arbitrarily;\;
let $S$ be the list of the manipulators' votes\;
\uIf{$(p$ not a unique winner of $(C, V\cup S)$ 
with weights $W_V\cup W_S)$}{return ``YES'';\;}
}
return ``NO'';\;
\caption{Algorithm for Bucklin-\textsc{DCWM}}
\label{alg:bv-dcwm}
\end{algorithm}

Before proving the runtime and correctness of the above algorithm, we
state the following useful lemma, which is easily seen to hold.

\begin{lemma}
\label{lem:bv-dcwm}
Let $(C,V)$ be a weighted Bucklin election with weights $W$ and $c,p\in C$. 
Then the following holds. 

\begin{enumerate}
\item Assume that $c$ is not a winner in $(C,V)$ and that the votes in
  $V$ are changed in a way such that only the position of $c$ is made
  worse.  Then $c$ is still not a winner.

\item Assume that $c$ is a winner of the election and that the votes
  in $V$ are changed in a way such that only the position of $c$ is
  improved. Then $c$ remains a winner.

\item Assume that $c$ is a winner of the election and that $p$ is not
  a winner. If in some votes the positions of candidates are swapped
  without changing the positions of $c$ and $p$, then
  $p$ is still not a winner.
\end{enumerate}
\end{lemma}

We now analyze Algorithm~\ref{alg:bv-dcwm} for Bucklin-\textsc{DCWM}.

\begin{theorem}
\label{thm:bv-dcwm-p}
Algorithm~\ref{alg:bv-dcwm} has a runtime in  
$\mathcal{O}(m^{2} (n+\|W_S\|))$
and decides  Bucklin-\textsc{DCWM}.
\end{theorem}

\begin{proofs}
  We begin with analyzing the runtime.  Obviously, the algorithm
  always terminates and the input size is in
  $\mathcal{O}(\underbrace{m}_{\|C\|}+\underbrace{n m}_{\|V\|}+
  \underbrace{n}_{\|W_{V}\|}+\|W_{S}\|+\underbrace{1}_{\|p\|}) =
  \mathcal{O}(n m+\|W_{S}\|)$.

  The most costly part of the algorithm is the for-loop.  To construct
  the manipulators' votes, $\mathcal{O}(\|W_{S}\| m)$ steps are
  needed.  The winner-determination procedure for Bucklin can be
  implemented with a runtime of $\mathcal{O}(n m)$, so the
  if-statement in line 8 can be computed in time $\mathcal{O}(m
  (n+\|W_{S}\|)$.  Thus, the whole for-loop runs in time
  $\mathcal{O}(m^{2} (n+\|W_{S}\|))$.

  To prove the correctnes of the algorithm, we show that it gives the
  output ``YES'' if and only if $(C,V,W_{V},W_{S},p)\in$
  Bucklin-\textsc{DCWM}.

  From left to right: If the algorithm outputs ``YES'' in line 2 then
  we have $\sum_{w\in W_{S}}w>\sum_{w\in W_{V}}w$, i.e., the sum of
  the manipulators' weights is greater than the sum of the weights of
  the nonmanipulative voters.  In this case, any of the candidates
  $c\neq p$ can be made the unique level~$1$ Bucklin winner in $(C,V\cup
  S)$ by putting $c$ on the first position in all the manipulators'
  votes and filling the remaining positions arbitrarily.  Hence,
  $(C,V,W_{V},W_{S},p)\in$ Bucklin-\textsc{DCWM}.
  If the algorithm outputs ``YES'' in line 9, the manipulators' votes
  have been constructed such that $p$ is not a unique winner in
  $(C,V\cup S)$.  Thus, we have that $(C,V,W_{V},W_{S},p)$ is a
  yes-instance of Bucklin-\textsc{DCWM}.

  From right to left: Assume that $(C,V,W_{V},W_{S},p)\in$
  Bucklin-\textsc{DCWM}.  If $\sum_{w\in W_{S}}w>\sum_{w\in W_{V}}w$,
  then the algorithm correctly outputs ``YES.''  Otherwise, the
  following holds: Since the given instance is a yes-instance of
  Bucklin-\textsc{DCWM}, the votes of the manipulators in $S$ can be
  set such that $p$ is not a winner of the election $(C,V\cup S)$.  We
  know from Lemma~\ref{lem:bv-dcwm} that successively swapping $p$
  with her neighbor until $p$ is on the last position in all votes in
  $S$ does not change the fact that $p$ is not a winner in $(C,V\cup
  S')$ (where $S'$ are the new manipulative votes with $p$ on the last
  position).  Assume that $c\in C-\{p\}$ is a winner in $(C,V\cup S)$.
  Then swap her position successively with her neighbor in the votes
  in $S'$ until $c$ is on the first position in all manipulative
  votes.  Let $S''$ denote the accordingly changed list of
  manipulative votes.  Again, from Lemma~\ref{lem:bv-dcwm} we know
  that $c$ still wins in $(C,V\cup S'')$.  Let $S'''$ be the list of
  manipulative votes that the algorithm constructs.  We can transform
  $S''$ into $S'''$ by swapping the corresponding candidates
  $c',c''\in C-\{c,p\}$ accordingly.  Since the positions of $c$ and
  $p$ remain unchanged, we have with Lemma~\ref{lem:bv-dcwm} that
  $p$ is still not a winner in $(C,V\cup S''')$.
  Thus, the algorithm outputs ``YES'' in line~$9$.~\end{proofs}

%

\section{Bribery in Bucklin and Fallback Voting}
\label{sec:bribery}

%

\subsection{Definition of Bribery Problems and Overview of Results}
\label{subsec:stdrd-brib}

We begin with defining the standard bribery scenarios proposed by
Faliszewski et al.~\cite{fal-hem-hem:j:bribery} (see
also~\cite{fal-hem-hem-rot:j:llull-copeland-full-techreport}) that
will be applied here to fallback and Bucklin elections.  Let
$\mathcal{E}$ be a given election system.

\EP{$\mathcal{E}$-Constructive Unweighted Bribery ($\mathcal{E}$-CUB)}
{An $\mathcal{E}$ election $(C,V)$, a designated candidate $p$, and a
nonnegative integer $k$.}
{Is it possible to make $p$ the unique $\mathcal{E}$ winner by
  changing the votes of at most $k$ voters?}

This basic bribery scenario can be extended by either considering
voters with different weights, or allowing that each voter has a
different price for changing her vote, or both.  These three scenarios
are formally defined by the following problems:

\EP{ $\mathcal{E}$-Constructive Weighted Bribery ($\mathcal{E}$-CWB)}
{An $\mathcal{E}$ election $(C,V)$ with each voter $v_i \in V$ having
  a nonnegative integer weight~$w_i$, a designated candidate~$p$, and
  a positive integer~$k$.}
{Is it possible to make $p$ the unique $\mathcal{E}$ winner by
  changing the votes of at most $k$ voters?}

\EP{$\mathcal{E}$-Constructive Unweighted Priced Bribery
  ($\mathcal{E}$-CUB-\$)}
{An $\mathcal{E}$ election $(C,V)$ with each voter $v_i \in V$ having
  a nonnegative integer price~$\pi_i$, $1 \leq i \leq n$, a designated
  candidate~$p$, and a positive integer~$k$.}
{Is there a set $B\seq \{1,\ldots,n\}$ such that $\sum\limits_{i\in
    B}\pi_i \leq k$ and the voters $v_i$ with $i \in B$ can be bribed
  so that $p$ is the unique $\mathcal{E}$ winner in the resulting election?}


\EP{$\mathcal{E}$-Constructive Weighted Priced Bribery ($\mathcal{E}$-CWB-\$)}
{An $\mathcal{E}$ election $(C,V)$ with each voter $v_i \in V$ having
  nonnegative integer weight $w_i$ and price~$\pi_i$, $1 \leq i \leq n$, 
  a designated candidate~$p$, and a positive integer~$k$.}
{Is there a set $B\seq \{1,\ldots,n\}$ such that $\sum\limits_{i\in
    B}\pi_i \leq k$ and the voters $v_i$ with $i \in B$ can be bribed
  so that $p$ is the unique $\mathcal{E}$ winner in the resulting
  election?}

By changing the question in the above four problems to ask whether $p$
can be prevented from being a unique winner of the election by bribing
some of the voters, we obtain the destructive variants of these
bribery scenarios, and we denote the corresponding problems by
$\mathcal{E}$-\textsc{DUB}, $\mathcal{E}$-\textsc{DWB},
$\mathcal{E}$-\textsc{DUB}-\$, and $\mathcal{E}$-\textsc{DWB}-\$.  The
problems related to the general bribery scenarios without explicitly
specifying the constructive or destructive case are denoted by
$\mathcal{E}$-\textsc{UB}, $\mathcal{E}$-\textsc{WB},
$\mathcal{E}$-\textsc{UB}-\$, and $\mathcal{E}$-\textsc{WB}-\$.

%
%

\begin{table} 
\caption{Overview of results for bribery in Bucklin and fallback voting}
\label{tab:overview-bribery}
 \centering
 \begin{tabular}{@{} l l l l l @{}}
\toprule
				& \multicolumn{2}{c}{Bucklin voting}
                                & \multicolumn{2}{c}{fallback voting} \\
                                & complexity & reference
                                & complexity & reference \\
\cmidrule(r){1-1} \cmidrule(rl){2-3} \cmidrule(l){4-5} 
$\mathcal{E}$-\textsc{CUB}	& $\np$-complete & Thm.~\ref{thm:bv-np-hard-CUB}	
				& $\np$-complete & Thm.~\ref{thm:fv-np-hard-CUB} 	\\
$\mathcal{E}$-\textsc{DUB}	& $\p$ & Cor.~\ref{cor:bv-p-dub}		
				& $\p$ & Thm.~\ref{thm:fv-p-dwb-dub}	\\
$\mathcal{E}$-\textsc{CUB-\$}	& $\np$-complete & Cor.~\ref{cor:bv-np-hard-brib}								
				& $\np$-complete & Cor.~\ref{cor:fv-np-hard-brib}	\\
$\mathcal{E}$-\textsc{DUB-\$}	& $\p$ & Thm.~\ref{thm:bv-p-dwb} 
				& $\p$ & Thm.~\ref{thm:fv-p-dwb-dub}	\\
$\mathcal{E}$-\textsc{CWB}	& $\np$-complete & Cor.~\ref{cor:bv-np-hard-brib} 
				& $\np$-complete & Cor.~\ref{cor:fv-np-hard-brib}	\\
$\mathcal{E}$-\textsc{DWB}	& $\p$ & Thm.~\ref{thm:bv-p-dwb} 
				& $\p$ & Thm.~\ref{thm:fv-p-dwb-dub}	\\
$\mathcal{E}$-\textsc{CWB-\$}	& $\np$-complete & Cor.~\ref{cor:bv-np-hard-brib}
				& $\np$-complete & Cor.~\ref{cor:fv-np-hard-brib}	\\
$\mathcal{E}$-\textsc{DWB-\$}	& $\np$-complete & Thm.~\ref{thm:bv-np-hard-DWB-dollar}
				& $\np$-complete & Thm.~\ref{thm:fv-np-hard-DWB-dollar}	\\
\bottomrule	 
\end{tabular}
\end{table}

Table~\ref{tab:overview-bribery} gives an overview of our complexity
results for bribery in Bucklin and fallback voting.

\subsection{Results for Bribery}
\label{subsec:stdrd-brib-results}

We start with the constructive cases of the standard bribery scenarios. 
\begin{theorem}
\label{thm:bv-np-hard-CUB}
\textsc{CUB} is $\np$-complete for Bucklin voting.
\end{theorem}
 
 \begin{proofs}
   Membership of Bucklin-\textsc{CUB} in $\np$ is obvious.
   We show $\np$-hardness by a reduction from \textsc{Exact Cover by
     Three-Sets (X3C)}: Given a set $B = \{b_1, b_2, \ldots ,
   b_{3m}\}$, $m\geq 1$, and a collection $\mathcal{S} = \{S_1, S_2,
   \ldots , S_n\}$ of subsets $S_i \seq B$ with $\|S_i\| = 3$ for
   each~$i$, $1 \leq i \leq n$, is there a subcollection $\mathcal{S}'
   \seq \mathcal{S}$ such that each element of $B$ occurs in exactly
   one set in~$\mathcal{S}'$?  \textsc{X3C} is a well-known
   $\np$-complete problem (see, e.g., \cite{gar-joh:b:int}).

   Let $(B,\mathcal{S})$ be an instance of \textsc{X3C} with
   $B=\{b_1,b_2,\ldots,b_{3m}\}$ and
   $\mathcal{S}=\{S_1,S_2,\ldots,S_n\}$.  Without loss of generality,
   we may assume that $n\geq 2m$.  We construct a Bucklin-\textsc{CUB}
   instance $((C,V),p,k)$, where $(C,V)$ is a Bucklin election with
   the candidates $C=B\cup\{c,d\}\cup G \cup\{p\}$, $p$ is the
   designated candidate, and $k=m$.  $G$ is a set of ``padding
   candidates,'' which are used to ensure that certain candidates do
   not gain points up to a certain level.  Padding candidates are
   positioned in the votes such that, up to a certain level, they do
   not gain enough points (e.g., at most one) to be relevant for the
   central argumentation of the proof.  Thus, their scores are not
   listed in tables giving the scores of the relevant candidates.
 
   For every $b_j\in B$, define $\ell_j$ to be the number of sets
   $S_i\in \mathcal{S}$ candidate $b_j$ is contained in.  $V$ consists
   of the following $2n$ voters (i.e., a strict majority is reached
   with $n+1$ votes):
 \begin{itemize}
 \item The first voter group consists of $n$ voters. For each $i$,
   $1\leq i \leq n$, we have one voter of the form
\[
c > d > S_i > G_1 > \{C-(\{c,d\}\cup S_i \cup G_1)\},
\]
where $G_1\seq G$ is a set of 
$3m-3$
padding candidates.  When a set
$X$ of candidates is giving in such a ranking, the order of the
candidates from $X$ can be fixed arbitrarily in this ranking.
  
\item The second voter group consists of $n$ voters as well.  We will
  present the preferences level by level from the first to the
  $(3m+2)$nd 
  position in Table~\ref{tab:bv-np-hard-CUB-(C,V)}.
 
\begin{table}[h!t]
\caption{Construction of Bucklin election $(C,V)$ in the proof of
 Theorem~\ref{thm:bv-np-hard-CUB}}
\label{tab:bv-np-hard-CUB-(C,V)}
 \centering
 \begin{tabular}{@{} c  c c@{}}
   \toprule
   position 		& \# voters 	& candidate \\
    \cmidrule(r){1-1} \cmidrule(rl){2-2} \cmidrule(l){3-3}
  \multirow{3}{*}{$1$} 	& $m$ 		& $c$\\
			& $m$		& $d$ \\
			& $n-2m$	& $g_k$\\
\cmidrule(r){1-1} \cmidrule(rl){2-2} \cmidrule(l){3-3}
 \multirow{2}{*}{$2$} 	& $n+1-\ell_1$ 	& $b_1$\\
			& $\ell_1-1$	& from $G_2$ \\
    \cmidrule(r){1-1} \cmidrule(rl){2-2} \cmidrule(l){3-3}
 \multirow{2}{*}{$3$} 	& $n+1-\ell_2$ 	& $b_2$\\
			& $\ell_2-1$	& from $G_2$ \\
    \cmidrule(r){1-1} \cmidrule(rl){2-2} \cmidrule(l){3-3}
		$\vdots$& $\vdots$	& $\vdots$\\
    \cmidrule(r){1-1} \cmidrule(rl){2-2} \cmidrule(l){3-3}
  \multirow{2}{*}{$3m+1$} 
 & $n+1-\ell_{3n}$ 	& $b_{3m}$\\
			& $\ell_{3n}-1$	& from $G_2$ \\
    \cmidrule(r){1-1} \cmidrule(rl){2-2} \cmidrule(l){3-3}
  \multirow{2}{*}{$3m+2$} 	& $n-m+1$ 		& $p$\\
			& $m-1$		& from $G_2$ \\
			\bottomrule
 \end{tabular}
\end{table}

\end{itemize}

Note that the padding candidates in $G$ are positioned in the votes
such that every $g_k\in G$ gains at most one point up to 
level~$3m+2$.
Table~\ref{tab:bv-np-hard-CUB-scores-(C,V)-original} shows the scores
of the relevant candidates in $(C,V)$ (namely, $c$, $d$, $p$, and each
$b_j\in B$) for the relevant levels 
(namely, $1$, $2$, $3m$, $3m+1$,
and $3m+2$)
and, in particular, that $c$ is the unique level~$1$ Bucklin
winner in $(C,V)$.

\begin{table} 
\captionsetup{font=small}
\caption{Level~$i$ scores for 
$i \in \{1,2,3m,3m+1,3m+2\}$ 
and the candidates in $C-G$}
\label{tab:bv-np-hard-CUB-scores-(C,V)}
\centering
\subfigure[Original election $(C,V)$]{
\label{tab:bv-np-hard-CUB-scores-(C,V)-original}
\begin{tabular}{@{} l c c c c @{}}
\toprule
			& $b_i\in B$ 	& $c$	& $d$	& $p$	\\
\cmidrule(r){1-1} \cmidrule(rl){2-2} \cmidrule(rl){3-3} \cmidrule(l){4-4} \cmidrule(l){5-5}
$\scorelev{1}$		& $0$ 		& $n+m$ & $m$	& $0$   	\\
$\scorelev{2}$ 		& $\leq n+1$	& $n+m$	& $m+n$	& $0$		\\
$\scorelev{3m}$ 	& $\leq n+1$ 	& $n+m$ & $m+n$	& $0$       	\\
$\scorelev{3m+1}$ 	& $\leq n+1$ 	& $n+m$ & $m+n$	& $0$		\\
$\scorelev{3m+2}$ 	& $n+1$ 	& $n+m$ & $m+n$	& $n-m+1$	\\
\bottomrule	 
\end{tabular}
}
\qquad
\subtable[Modified election $(C,V')$]{
\label{tab:bv-np-hard-CUB-scores-(C,V)-modified}
\begin{tabular}{@{} l c c c c @{}}
\toprule
			& $b_i\in B$ 	& $c$	& $d$	& $p$	\\
\cmidrule(r){1-1} \cmidrule(rl){2-2} \cmidrule(rl){3-3} \cmidrule(l){4-4} \cmidrule(l){5-5}
$\scorelev{1}$		& $0$ 		& $n$ 	& $m$	& $m$   	\\
$\scorelev{2}$ 		& $\leq n$	& $n$	& $n$	& $m$		\\
$\scorelev{3m}$ 	& $\leq n$ 	& $n$ 	& $n$	& $m$       	\\
$\scorelev{3m+1}$ 	& $\leq n$ 	& $n$ 	& $n$	& $m$		\\
$\scorelev{3m+2}$ 	& $\leq n$ 	& $n$ 	& $n$	& $n+1$	\\
\bottomrule	 
\end{tabular}
}
\end{table}
 
We claim that $\mathcal{S}$ has an exact cover $\mathcal{S}'$ for $B$
if and only if $p$ can be made the unique Bucklin winner by changing
at most $m$ votes in $V$.
 
From left to right: Let $\mathcal{S}'$ be an exact cover for $B$ and
let $I\seq\{1,\ldots,n\}$ be the set of indices of the $m$ elements in
$\mathcal{S}'$.  To make $p$ the unique Bucklin winner, we only have
to change votes in the first voter group: For each $i\in I$, change
the corresponding vote
\[
c > d > S_i > G_1 > \{C-(\{c,d\}\cup S_i \cup G_1)\}
\]
to 
\[
p > G_1 > g'_1 > g'_2 > g'_3 > g'_4 > \{C-(\{g'_1,g'_2,g'_3,g'_4,p\}\cup G_1)\},
\]
where each $g'_j$, $1 \leq j \leq 4$, is from $G$ but not in~$G_1$.

 
With these new votes, $c$ and $d$ both lose $m$ points on the first
two levels from the first voter group and $p$ gains $m$ points on the
first level.  Every candidate $b_i\in B$ loses exactly one point on
one of the levels~$3$, $4$, or~$5$.  The scores in the resulting
election $(C,V')$ are shown in
Table~\ref{tab:bv-np-hard-CUB-scores-(C,V)-modified}.  As one can see,
$p$ is the first candidate to reach a strict majority of $n+1$ votes
(namely, on level~$3m+2$) 
and is thus the unique 
level~$3m+2$
Bucklin
winner in the new election.
 
From right to left: Assume that $p$ is the unique Bucklin winner of the
election $(C,V')$, where $V'$ is the new voter list containing the $m$
changed votes.  Since only $m$ votes can be changed and $p$ did not
score any points prior to 
level~$3m+2$
in the original election, $p$
has to be a 
level~$3m+2$
Bucklin winner in $(C,V')$.  Candidates $c$ and
$d$ originally reach the majority threshold already on, respectively,
the first and the second level, so all votes that can be changed must
place $c$ and $d$ on the first two positions.  The only votes doing so
are those in the first voter group.  Finally, to prevent the
candidates in $B$ from reaching a strict majority on 
level~$3m+2$,
each of the $3m$ candidates has to lose at least one point by changing
at most $m$ votes. This, again, can only be done by changing votes
from the first voter group and there has to be an exact cover
$\mathcal{S}'$ for $B$ whose corresponding voters from the first voter
group have to be changed.~\end{proofs}

The following 
corollary follows
immediately from
Theorem~\ref{thm:bv-np-hard-CUB}.
 
\begin{corollary}
\label{cor:bv-np-hard-brib}
  In Bucklin elections, \textsc{CWB}, \textsc{CUB}-\$, and
  \textsc{CWB}-\$ are $\np$-complete.
\end{corollary}

Based on the corresponding proof for approval voting that 
is due to Faliszewski et al.~\cite{fal-hem-hem:j:bribery}, we can show 
$\np$-completeness for unweighted bribery in fallback elections
as well. 

\begin{theorem}
\label{thm:fv-np-hard-CUB}
\textsc{CUB} is $\np$-complete for fallback voting.
\end{theorem}

\begin{proofs}
 Fallback-\textsc{CUB} obviously is in $\np$. To show $\np$-hardness, we 
 give a reduction from \textsc{X3C}. Let $(B,\mathcal{S})$ be an instance 
 of \textsc{X3C} with   $B=\{b_1,b_2,\ldots,b_{3m}\}$ and
   $\mathcal{S}=\{S_1,S_2,\ldots,S_n\}$. 
 We define the fallback election $(C,V)$ with the candidate set 
 $C=B \cup E \cup \{p\}$, where $p$ is the designated candidate and $E$ 
 is a set of $n+m$ padding candidates. 
 For every $j\in\{1,\ldots,3m\}$, we define $\ell_j$ as the number of 
 subsets $S_i\in \mathcal{S}$ candidate $b_j\in B$ is contained in. 
 Using this notation,
 we define the subsets $B_i=\{b_j\in B \mid i\leq n-\ell_j\}$ 
 for $i\in \{1,\ldots,n\}$. 
 $V$ consists of the $4n$ voters whose preferences are given in
 Table~\ref{tab:fv-np-hard-CUB}. 
 
\begin{table}[h!t]
\caption{Construction of fallback election $(C,V)$ in the proof of
 Theorem~\ref{thm:fv-np-hard-CUB}}
\label{tab:fv-np-hard-CUB}
\centering
\begin{tabular}{@{} l l c l  @{}}
\toprule
	\#		& For each $\ldots$	& number of votes	&ranking of candidates	\\
\cmidrule(r){1-1} \cmidrule(rl){2-2} \cmidrule(rl){3-3} \cmidrule(l){4-4} 
$1$	& $i\in\{1,\ldots,n\}$ 	& $1$ 	& $S_i\, \mid \, (B-S_i)\cup E\cup\{p\} $	  	\\
$2$ 	& $i\in\{1,\ldots,n\}$ 	& $1$ 	& $B_i\, \mid \, (B-B_i)\cup E\cup\{p\} $	  	\\
$3$ 	&  			& $n-m$ & $p\, \mid \, B\cup E$	      	\\
$4$ 	& $\ell\in\{1,\ldots,n+m\}$ & $1$ & $e_\ell\, \mid \, B\cup(E-\{e_\ell\}\cup\{p\} $\\
\bottomrule	 
\end{tabular}
\end{table}

In this election, we have that $\scoreof{p}=n-m$, $\scoreof{b_j}=n$ for all 
$j\in \{1,\ldots,3m\}$, and $\scoreof{e_\ell}=1$ for all $\ell\in\{1,\ldots,n+m\}$. 
Since no candidate reaches a strict majority (at least
$2n+1$ points), all candidates $b_j \in B$ are fallback winners of this
election.

%
We claim that $\mathcal{S}$ has an exact cover $\mathcal{S}'$ for $B$
if and only if $p$ can be made the unique fallback winner by bribing at most $m$ voters.

From left to right: Suppose that $\mathcal{S}$ has an exact cover $\mathcal{S}'$
for $B$. We change the vote of those voters in the first voter group where
$S_i \in \mathcal{S}'$ from $S_i \, \mid \, (B-S_i) \cup E \cup \{p\}$ to 
$p \, \mid \,B \cup E$. In the resulting election $(C,V')$ only the scores 
of the candidates in $B$ and the score of $p$ change: $p$ gains $m$ points, 
whereas each $b_j\in B$ loses exactly one point. Thus, with an overall score of 
$n$, candidate $p$ is the unique fallback winner of the resulting election. 

From right to left: Suppose that $p$ can be made the unique  fallback winner by
changing at most $m$ votes in $V$. That means that $p$ can gain at
most $m$ points, so the maximum overall score that $p$ can reach is
$n$. Since each $b_j \in B$ has an overall score of $n$, every
candidate in $B$ has to lose at least one point by changing at most
$m$ votes (otherwise, there would be at least one candidate in $B$ that
ties with $p$).  This is possible only if in $m$ votes of the first voter
group the candidates in $S_i$ are removed from the approval strategy such
that these $m$ sets $S_i$ form an exact cover for~$B$.~\end{proofs}

This result immediately implies $\np$-hardness for 
the remaining constructive bribery scenarios in fallback 
elections as well.

\begin{corollary}
\label{cor:fv-np-hard-brib}
  In fallback elections, \textsc{CWB}, \textsc{CUB}-\$, and
  \textsc{CWB}-\$ are $\np$-complete.
\end{corollary}
%

We now turn to the destructive cases.  The following result
generalizes a result due to Xia~\cite{xia:c:margin-of-victory} who
showed that \textsc{DUB} is in $\p$ for simplified Bucklin voting.

\begin{theorem}
\label{thm:bv-p-dwb}
  In Bucklin elections, \textsc{DWB} and \textsc{DUB-\$} are in $\p$.
\end{theorem}

\begin{proofs} 
  Both problems, Bucklin-\textsc{DWB} and Bucklin-\textsc{DUB-\$}, can
  be solved by deterministic polynomial-time algorithms that use
  Algorithm~\ref{alg:bv-dcwm}, which was designed in
  Section~\ref{sec:manipulation} to solve the destructive coalitional
  weighted manipulation problem for Bucklin elections,
  Bucklin-\textsc{DCWM}.  The main difference between a bribery and a
  manipulation instance is that in the latter only the preferences of
  the manipulators have to be found, whereas in the former both the
  votes that will be bribed and the new preferences for these voters
  have to be found.  If we have the set of votes we want to change, we
  can use the algorithm for the manipulation problem to construct the
  preferences.  Thus, for the runtime of the algorithm the
  determination of these voter sets is crucial, and we show that in
  Bucklin elections the number of voter sets whose modification might
  actually lead to a successful bribery is bounded by a polynomial in
  both the number of voters and the number of candidates.

  Consider Algorithm~\ref{alg:bv-dwb} and a given input
  $(C,V,W_V,p,k)$ to it.  In particular, $p$ is the designated
  candidate that we want to prevent from winning and assume that we
  have a yes-instance, i.e., our bribery action is successful.  We
  denote by $(C,V'')$ the election resulting from $(C,V)$ where the
  $k$ votes that can be changed have already been changed.  Then there
  is a candidate $c\in C-\{p\}$ that reaches a strict majority on
  level~$i$, and it holds that $\scoresublevel{(C,V'')}{i}{c}\geq
  \scoresublevel{(C,V'')}{i}{p}$, which means that $p$ is not a unique
  winner in $(C,V'')$.  To reach that, for each $i<m$,
  there are only five types of
  preferences that might have been changed in~$V$, and they can be
  grouped into the following subsets $T_{i,j}\seq V$, $1 \leq j\leq 5$:
\begin{description}
 \item[$T_{i,1}$:] $p$ is among the top $i-1$ position and $c$ is among the 
 top $i$ positions (when changing: $p$ loses points, $c$ does neither 
 lose nor win points up to level $i$).
 \item[$T_{i,2}$:]  $p$ is among the top $i-1$ position and $c$ is not among the 
 top $i$ positions (when changing: $p$ loses points, $c$ wins points
 up to level $i$).
 \item[$T_{i,3}$:]  $p$ is on position $i$ and $c$ is among the 
 top $i$ positions (when changing: $p$ loses points, $c$ does neither 
 lose nor win points up to level $i$).
 \item[$T_{i,4}$:]  $p$ is on position $i$ and $c$ is not among the 
 top $i$ positions (when changing: $p$ loses points, $c$ wins 
 points up to level $i$).
 \item[$T_{i,5}$:]  both $p$ and $c$ are not among the 
 top $i$ positions (when changing: $p$ does neither 
 lose nor win points, $c$ wins points up to level $i$).
\end{description}

For a sublist of voters $V'\seq V$, denote their total weight
by~$W_V'$.  Algorithm~\ref{alg:bv-dwb} for Bucklin-\textsc{DWB} works
as follows.

\begin{algorithm}  
\small
\DontPrintSemicolon
\SetKwInOut{Input}{input}\SetKwInOut{Output}{output}
\Input{$C$ set of candidates\\ 
~$V$ list of voters\\
~$W_V$ list of weights of voters\\
~$k$ number of votes that may be changed\\
~$p$ designated candidate}
\Output{``YES'' if $(C,V,W_V,k,p)\in\text{Bucklin-\textsc{DWB}}$ \\
~``NO'' if $(C,V,W_V,k,p)\notin\text{Bucklin-DWB}$}
\BlankLine
let $A=\{(a_1,a_2,\ldots,a_5)\condition a_i\in\{0,1,\ldots,k\}\}$,
    $V'=\emptyset$;
\\
\ForEach{$c\in C-\{p\}$}
{ \ForEach{$i<m$}
  { 
  \ForEach{$(a_1,a_2,\ldots,a_5)\in A$}
    {
    \uIf{$\sum_{\ell=1}^{5}a_\ell\leq k$}
      {\ForEach{$j\in \{1,2,\ldots,5\}$}
	{add the $a_j$ heaviest votes in $T_{i,j}$ to $V'$;
	 }
	run Algorithm~\ref{alg:bv-dcwm} on input 
	$(C,V-V',W_{V-V'},W_{V'},p)$;\\
	\uIf{\textup{Bucklin-}\textsc{DCWM}$(C,V-V',W_{V-V'},W_{V'},p)$= ``YES''}
	  {
	  return ``YES'';
	  }
      }
    }
  }
}
return ``NO'';
\caption{Algorithm for Bucklin-\textsc{DWB}}
\label{alg:bv-dwb}
\end{algorithm}

It is easy to see that Algorithm~\ref{alg:bv-dwb} runs in
deterministic polynomial time: the two outer for-loops iterate up to
$m$ times, whereas the inner loop tests up to $k^5$ variations of the
vector $(a_1,a_2,\ldots,a_5)$.  Since $k\leq n$, we have that the
number of executions of Algorithm~\ref{alg:bv-dcwm} is in
$\mathcal{O}(m^2 n^5)$.

For the proof of correctnes, we show that given a bribery instance
$(C,V,W_V,k,p)$, the output of Algorithm~\ref{alg:bv-dwb} is ``YES''
if and only if $(C,V,W_V,k,p)\in$ Bucklin-\textsc{DWB}.

From left to right: If the algorithm returns ``YES'' in line~$10$,
then Algorithm~\ref{alg:bv-dcwm} could find a successful destructive
manipulation regarding $p$ for $k$ manipulators with total
weight~$W_{V'}$.  So $p$ is not a unique Bucklin winner in the election
$(C,V'')$, where $V''$ is the list of voters with $k$ changed votes.
That means that $(C,V,W_V,k,p)\in$ Bucklin-\textsc{DWB}.

From right to left: Assume that $(C,V,W_V,k,p)\in$
Bucklin-\textsc{DWB}.  Thus, there exists a set of $k$ voters $V'$
with total weight $W_{V'}$ such that changing these votes prevents $p$
from being a unique Bucklin winner in $(C,V'')$, where $V''$ is the new
voter list containing the $k$ changed votes.  We want to show that
such a $V''$ can always be transformed to the list of votes $V'$ that
is changed in Algorithm~\ref{alg:bv-dwb}.  From our assumptions it
follows that we have a candidate $c\in C-\{p\}$ and a level~$i<m$ such
that $c$ is a level~$i$ Bucklin winner that prevents $p$ from being a
unique winner.

Assume that in $V''$ are voters whose preferences were not in one of
the $T_i$ before the changes were made, i.e., votes were changed that
not necessarily needed to be changed to prevent $p$ from being the
only winner.  Undo these changes and change the same number of votes
in the lists $T_i$ that were not changed before.  We then have that
all changed votes are in one of the~$T_i$.

Since Bucklin is monotonic, we can always exchange votes with higher
weight with votes of lower weight (in one $T_i$) without risking that
$p$ would win due to this exchange.  So we know that we can transform
any given list of bribed votes to a list that the algorithm would
construct and this would still prevent $p$ from winning alone.  So if
there is a list of $k$ voters that can be successfully bribed to
prevent $p$ from being the unique winner, the algorithm will find it.

For the Bucklin-\textsc{DUB}-\$ problem the same algorithm can be
used. The only difference is that all weights have to be set to one,
the cheapest instead of the heaviest votes (i.e., those votes with the
least price instead of the greatest weight) are added to $V'$ in
line~$7$, and it has to be tested whether the sum of the chosen votes
does not exceed the budget.~\end{proofs}

From Theorem~\ref{thm:bv-p-dwb} we have the following corollary.

\begin{corollary}
\label{cor:bv-p-dub}
  In Bucklin elections, \textsc{DUB} is in~$\p$.
\end{corollary}

This algorithm can
be easily adapted for fallback elections.
Due to the fact that in fallback elections the voters do not have to
rank all candidates, it is possible that a candidate wins on
level~$m$.  So by making the following changes in
Algorithm~\ref{alg:bv-dcwm}:
\begin{itemize}
\item change ``$i < m$'' in line~$3$ to ``$i\leq m$,''
\item use the fallback analogue of Algorithm~\ref{alg:bv-dcwm} in
  line~$8$,\footnote{In Section~\ref{sec:weighted-mani},
  we refrained from explicitly stating the algorithm 
  for fallback-\textsc{DCWM} that is based on the following 
  simple idea: For every candidate $c\neq p$, try to make $c$ win
  by setting the manipulators' votes to $c\,\mid\,C-\{c\}$.
  If such a candidate can be found, $p$ has been successfully prevented from 
  being the unique winner of the election; otherwise, it is impossible to
  do so.}
 and
\item change
``Bucklin-\textsc{DCWM}'' in line~$9$ to ``Fallback-\textsc{DCWM},''
\end{itemize}
we can decide \textsc{DWB} for fallback elections as well.

\begin{theorem}
\label{thm:fv-p-dwb-dub}
In fallback elections, \textsc{DWB}, \textsc{DUB}, and \textsc{DUB}-\$
are in~$\p$.
\end{theorem}

It remains to show the complexity of the destructive variant of priced
bribery in weighted Bucklin and fallback elections.  
We begin with showing 
$\np$-hardness of this problem 
for Bucklin voting.

\begin{theorem}
\label{thm:bv-np-hard-DWB-dollar}
  Bucklin-\textsc{DWB-\$} is $\np$-complete.
\end{theorem}

\begin{proofs}
  That Bucklin-\textsc{DWB-\$} is in $\np$ is again easy to see.  We
  show $\np$-hardness by a reduction from \textsc{Partition}.  Let
  $(A,(a_1,\ldots,a_k))$ with $A = \{1,\ldots,k\}$ and $\sum_{i=1}^k
  a_i=2K$ be an instance of {\sc Partition}.  We construct the
  following Bucklin election $(C,V)$ with $C=\{c,p\}$ and $k$ votes
  in~$V$: For each $i\in \{1,\ldots,k\}$, we have one voter with weight
  $w_i=a_i$, price $\pi_i=a_i$, and preference $p>c$.

  The total weight in $(C,V)$ is $2K$. Let $K$ be the budget that may
  not be exceeded and let $p$ be the designated candidate.  Obviously,
  $p$ is the unique level~$1$ Bucklin winner in $(C,V)$.
 
  We claim that $(A,(a_1,\ldots,a_k))\in $ \textsc{Partition} if and
  only if $p$ can be prevented from being the unique Bucklin winner by
  changing votes in $V$ without exceeding the budget $K$.
 
  From left to right: Let $(A,(a_1,\ldots,a_k))\in $
  \textsc{Partition} with $A'\seq A$ such that $\sum_{i\in A'}a_i=K$.
  Change the votes of those voters with $w_i=a_i$ for $i \in A'$ from
  $p>c$ to $c>p$.  With these changes we have that on the first level,
  both $p$ and $c$ have exactly $K$ points, so no strict majority.  On
  the second level, both candidates have $2K$ points and thus are both
  level~$2$-Bucklin winners.  Hence, $p$ is not a unique Bucklin
  winner in the bribed election.
 
  From right to left: Assume that $p$ is not a unique Bucklin winner
  in the bribed election.  Since there are only two levels, either $c$
  is the unique level~$1$ winner, or $p$ and $c$ both win on the
  second level.  The price and weight of every voter is the same, so
  voters with a total weight of $K$ can be changed.  Candidate $c$ has
  $0$ points in the original election, so it is not possible to make
  $c$ a unique level~$1$ winner without exceeding the budget~$K$.  To
  prevent $p$ from remaining the unique winner on the first level, the
  budget has to be fully exhausted and votes with a total weight of
  $K$ must be changed from $p>c$ to $c>p$.  Thus, there is a subset
  $A'\seq A$ such that $\sum_{i\in A'}a_i=K$, so
  $(A,(a_1,\ldots,a_k))\in $ \textsc{Partition}.~\end{proofs}

%
Theorem~\ref{thm:fv-np-hard-DWB-dollar} states the corresponding 
result for fallback voting using a similar proof idea.

\begin{theorem}
\label{thm:fv-np-hard-DWB-dollar}
  Fallback-\textsc{DWB-\$} is $\np$-complete.
\end{theorem}

\begin{proofs}
Obviously, fallback-\textsc{DWB-\$} is in $\np$. $\np$-hardness is shown 
by a reduction from \textsc{Partition}. To this end, let 
  $(A,(a_1,\ldots,a_k))$ with $A = \{1,\ldots,k\}$ and $\sum_{i=1}^k
  a_i=2K$ be an instance of {\sc Partition}. We construct a 
  fallback election $(C,V)$ with the candidate set $C=\{c,p\}$ and 
  the designated candidate~$p$.  Let $V$ consist of $k$ 
  voters $v_1,\ldots,v_k$, each having preference $p\,\mid\,\{c\}$, 
  weight $w_i=a_i$, and price $\pi_i=a_i$.
  The total weight in $(C,V)$ is $2K$ and $p$ is the unique fallback 
  winner in this election.  Let the briber's budget be~$K$.
  
  We claim that $(A,(a_1,\ldots,a_k))\in $ \textsc{Partition} if and
  only if $p$ can be prevented from being the unique fallback winner by
  changing votes in $V$ without exceeding the budget $K$.
  
  From left to right: Assume that there is a set $A'\seq A$ such 
  that $\sum_{i\in A'}a_i=K$. We change the votes of each voter $v_i$ 
  with $i\in A'$ from $p\,\mid\,\{c\}$ to $c\,\mid\,\{p\}$. 
  Then both candidates, $p$ and~$c$, have an overall score of $K$ and are 
  both fallback winners of the resulting election. 
  
  From right to left: Assume that $p$ can be prevented from being the 
  unique fallback winner by changing votes in $V$ without exceeding 
  the budget $K$. This is possible only if candidate $c$ at least ties with 
  $p$ after the changes in the votes have been made. Since
  each voter's weight and price are the same, $c$ can gain at most $K$ points without 
  exceeding the budget $K$. To prevent $p$ from being the unique fallback winner, 
  $c$ has to gain at least $K$ points, so together with the budget restriction 
  $c$ has to gain exactly $K$ points.  Thus,
  there has to be a set  $A'\seq A$ such that $\sum_{i\in A'}a_i=K$.~\end{proofs}

\section{Campaign Management}

In the discussion so far, we have focused on bribery and manipulation
as means of attacking Bucklin and fallback elections. However, it is
also quite natural to consider bribery scenarios through the lenses of
running a political campaign. After all, in a successful campaign, the
candidates spend their effort (measured in terms of time, in terms of
financial cost of organizing promotional activities, and even in terms
of the difficulty of convincing particular voters) to change the minds
of the voters. Thus, formally, a campaign preceding an election can be
seen as exchanging some resources for voters' support. Formally, this
idea is very close to bribery (indeed, this view of campaign
management was first presented in a paper whose focus was on a bribery
problem~\cite{elk-fal-sli:c:swap-bribery}).

\subsection{Definitions and Overview of Results}

We start by discussing one of the most general campaign management
problems, namely the \textsc{Swap Bribery} problem introduced by Elkind et
al.~\cite{elk-fal-sli:c:swap-bribery}. This problem models a situation
where a campaign manager, who is interested in the victory of a given
candidate $p$, can organize meetings with specific voters (the
unweighted variant of the problem) or with groups of like-minded
voters (the weighted variant) and convince them to change their
preference orders. However, the difficulty (or, as we will say from
now on, the cost) of changing the voters' preference orders depends
both on the voter and on the extent of the change (for example, it
might be expensive to swap a voter's most preferred candidate with
this voter's least preferred one, but it might be very cheap to swap
the voter's two least preferred candidates). Formally, Elkind et
al.~\cite{elk-fal-sli:c:swap-bribery} define so-called
\emph{swap-bribery price functions} that for each voter and for each
pair of candidates give the cost of swapping these two candidates in
the voter's preference order (provided the candidates are adjacent in
this order).
\begin{definition}[Elkind et al.~\cite{elk-fal-sli:c:swap-bribery}]
\label{def:swap-price-fctn}
A \emph{swap-bribery price function} for voter $v_i$ is a function
$\pi_i:C\times C \rightarrow \mathbb{N}$ that specifies for any
ordered pair $(c_i,c_j)$ of candidates the price for changing $v_i$'s
preference order from $\cdots > c_i > c_j > \cdots$ to $\cdots > c_j >
c_i > \cdots$.  Only candidates that are adjacent in a vote can be
swapped.
\end{definition}
In the $\mathcal{E}$-\textsc{Constructive Swap Bribery} problem we ask if
there exists a sequence of swaps of adjacent candidates that lead to a
given candidate being a winner (note that the swaps are performed in
sequence; even if some candidates are not adjacent at first, they may
become adjacent in the course of performing the swaps and, then, can
be swapped themselves).\footnote{We mention that Elkind et
  al.~\cite{elk-fal-sli:c:swap-bribery} defined the problem for the
  nonunique-winner model. Here we adopt the unique-winner model to
  stay in sync with the rest of the paper. However, it will be easy to
  see that all the results from this section hold in the
  nonunique-winner model as well.}

\EP{$\mathcal{E}$-Constructive Unweighted Swap Bribery ($\mathcal{E}$-CUSB)}
{An $\mathcal{E}$-election $(C,V)$, where $V = (v_1, \ldots, v_n)$, a designated candidate $p$, a list
  $(\pi_1,\ldots,\pi_n)$ of swap bribery price functions, and a
  nonnegative integer~$k$.}
{Can $p$ be made the unique $\mathcal{E}$ winner of an election
  resulting from the input election
  by conducting a sequence of swaps of adjacent candidates in the
  voters' ballots such that the total cost of the swaps does not
  exceed the budget~$k$?}

We define the weighted variant of the problem, $\mathcal{E}$-CWSB, in
the standard way (as far as we can tell, the weighted variant of the
problem has not been studied before). However, it will soon become clear
why the weighted variant is not particularly interesting and so we
omit the easy modification of the definition.
%
%
We also define the destructive variants of the swap bribery problems
($\mathcal{E}$-DUSB-\$ and $\mathcal{E}$-DWSB-\$) in the usual way, by
changing the question to ask whether $p$ can be prevented from being a
unique winner.

Swap bribery is a very difficult problem---it is $\np$-complete for
almost all natural voting rules (and, in particular, in the next
section we will see a very strong hardness result for the Bucklin and
fallback rules). Thus Elkind et al.~\cite{elk-fal-sli:c:swap-bribery}
defined its much-simplified variant, shift-bribery, where every swap
has to involve the designated candidate $p$ (that is, the designated
candidate can be ``shifted'' forward in selected votes).  The
complexity of this problem was studied for a number of voting
rules~\cite{elk-fal-sli:c:swap-bribery,elk-fal:c:shift-bribery,dor-sch:j:parameterized-swap-bribery},
including Bucklin and
fallback voting~\cite{sch-fal-elk:c:campaign-management-under-approval-driven-voting}.
Interestingly, even though we will see strong hardness results for
swap bribery under Bucklin and fallback, shift bribery for these rules
is in $\p$.

The definitions of swap bribery and shift bribery are very natural for
voting rules where each voter ranks all the candidates; for the case
of fallback, where the ballots consist of the approved part (where the
candidates are ranked) and of the disapproved part (where the
candidates are not ranked), we need to extend the definitions.  In our
approach, we define swap bribery under fallback to allow the swaps
within the approved parts of the votes only. Naturally, one could also
define costs for including given disapproved candidates in the
approved part and, indeed, Elkind et
al.~\cite{elk-fal-sli:c:swap-bribery} did so for SP-AV (SP-AV is a
variant of the approval system).\footnote{Like fallback voting, SP-AV
  is a hybrid variant of approval voting.  It has been introduced by
  Brams and Sanver~\cite{bra-san:j:critical-strategies-under-approval}
  and slightly modified by Erd\'{e}lyi et
  al.~\cite{erd-now-rot:j:sp-av} to cope with certain control actions
  (see also the chapter by Baumeister et
  al.~\cite{bau-erd-hem-hem-rot:b:computational-aspects-of-approval-voting}
  for a through discussion of this voting system).}  However,
following Schlotter et
al.~\cite{sch-fal-elk:c:campaign-management-under-approval-driven-voting},
we believe that it is more informative to study the complexity of
modifying the rankings within the approved parts and the complexity of
modifying the sets of approved candidates separately.

Regarding the latter type of problems, Schlotter et
al.~\cite{sch-fal-elk:c:campaign-management-under-approval-driven-voting}
defined the support bribery problem for the fallback rule (and other hybrid
rules), where each voter has a complete preference order over the
whole set of candidates, but also has an approval threshold, a number
of top candidates that this voter approves of. For each voter we have
a price function that gives the cost of increasing/decreasing the
approval threshold; the goal is to change the voters' approval thresholds
in such a way as to ensure the victory of a given candidate. Schlotter
et
al.~\cite{sch-fal-elk:c:campaign-management-under-approval-driven-voting}
show that this problem is $\np$-complete for fallback.\footnote{They
  also show that the problem is hard in the sense of
  parametrized complexity for two
  natural parameters describing the extent of change to the approval
  thresholds. Interestingly, they show the problem to be
  fixed-parameter tractable if the
  thresholds can either only increase or only decrease.}
However, in our model the disapproved candidates are not ranked and,
thus, it is much more natural to study the extension bribery problem
introduced by Baumeister et
al.~\cite{bau-fal-lan-rot:c:campaigns-for-lazy-voters}. The idea of
extension bribery is to capture very non-invasive campaign actions,
where we try to convince some voters to include the designated
candidate at the end of the ranking of approved candidates.

\begin{definition}[Baumeister et
  al.~\cite{bau-fal-lan-rot:c:campaigns-for-lazy-voters}] The
  \emph{extension bribery price function} $\delta_{i} : \mathbb{N}
  \rightarrow \mathbb{N}$ of a voter $v_i$ defines the price for
  extending the approved part of $v_i$'s vote with a given number of
  so-far-disapproved candidates (these new candidates are ranked below
  the previously-approved candidates, but among themselves are ranked
  as the briber requests).
\end{definition}

We define the following related problem.

\EP{FV-Constructive Unweighted Extension Bribery (FV-CUEB)} {A
  fallback election $(C,V)$, where $V = (v_1, \ldots, v_n)$, a
  designated candidate $p$, a list $(\delta_1,\ldots,\delta_n)$ of
  extension bribery price functions, and a nonnegative integer~$k$.}
{Can $p$ be made the unique fallback winner by extending the approved
  parts of the the voters' ballots without exceeding the budget~$k$?}

Again, the weighted variant ($\mathcal{E}$-CWEB) is defined in the
natural way and so are the destructive variants ($\mathcal{E}$-DUEB
and $\mathcal{E}$-DWEB).

Table~\ref{tab:overview-swap-extension-bribery} summarizes the results
of this section.

\begin{table} 
  \caption{Overview of results for swap bribery and extension bribery in Bucklin and fallback voting}
\label{tab:overview-swap-extension-bribery}
 \centering
 \begin{tabular}{@{} l l l l l @{}}
\toprule
				& \multicolumn{2}{c}{Bucklin voting}
                                & \multicolumn{2}{c}{fallback voting} \\
                                & complexity & reference
                                & complexity & reference \\
\cmidrule(r){1-1} \cmidrule(rl){2-3} \cmidrule(l){4-5} 
$\mathcal{E}$-\textsc{CUSB}	& $\np$-complete & Thm.~\ref{thm:bv-cusb}	
				& $\np$-complete & Cor.~\ref{thm:bv-fv-usb} 	\\
$\mathcal{E}$-\textsc{DUSB}	& $\np$-complete & Cor.~\ref{thm:bv-fv-usb}		
				& $\np$-complete & Cor.~\ref{thm:bv-fv-usb}	\\
$\mathcal{E}$-\textsc{CWSB}	& $\np$-complete & Thm.~\ref{thm:weighted-swap}
				& $\np$-complete & Thm.~\ref{thm:weighted-swap}	\\
$\mathcal{E}$-\textsc{DWSB}	& $\np$-complete & Thm.~\ref{thm:weighted-swap} 
				& $\np$-complete & Thm.~\ref{thm:weighted-swap}	\\
$\mathcal{E}$-\textsc{CUEB}	& -- &  
				& $\p$ & Thm.~\ref{thm:fv-cueb-dueb}	\\
$\mathcal{E}$-\textsc{DUEB}	& -- &  
				& $\p$ & Thm.~\ref{thm:fv-cueb-dueb}	\\
$\mathcal{E}$-\textsc{CWEB}	& -- & 
				& $\np$-complete & Thm.~\ref{thm:fv-w-p-micro}	\\
$\mathcal{E}$-\textsc{DWEB}	& -- & 
				& $\np$-complete & Thm.~\ref{thm:fv-w-p-micro}	\\
\bottomrule	 
\end{tabular}
\end{table}

\subsection{Results for Swap Bribery}

We start by quickly observing that weighted swap bribery is
$\np$-complete for both Bucklin and fallback rules. 

\begin{theorem}\label{thm:weighted-swap}
  \textsc{BV-CWSB}, \textsc{BV-DWSB}, \textsc{FV-CWSB}, and \textsc{FV-DWSB} are $\np$-complete.
\end{theorem}
\begin{proofs}
  The proof for Bucklin is a direct consequence of the fact that
  CWB-\$ is $\np$-complete for plurality, even for just two
  candidates~\cite{fal-hem-hem:j:bribery} (the result holds both for
  the uniqe-winner case and for the nonunique-winner case).  For two
  candidates, the Bucklin rule is identical to the plurality rule.
  Further, for two candidates CWB-\$ is, in essence, identical to CWSB
  (the only possible bribery is to swap the only
  two candidates), and the nonunique-winner variant of CWB-\$ is, in
  essence, identical to DWSB.

  For fallback, membership of the problems in $\np$ is clear, and
  $\np$-hardness follows by the same arguments as for Bucklin, by
  considering the setting where every voter approves of all
  candidates.~\end{proofs}

For the unweighted case, $\np$-completeness of \textsc{BV-CUSB}
follows immediately from the fact that the possible winner problem for
Bucklin is $\np$-complete (see the papers of Konczak and
Lang~\cite{kon-lan:c:incomplete-prefs}, for the definition of the
possible winner problem, and of Xia and
Conitzer~\cite{con-xia:j:possible-necessary-winners}, for the result
regarding Bucklin) and the fact that, for a given voting rule, the
possible winner problem reduces to the swap bribery
problem~\cite{elk-fal-sli:c:swap-bribery}. However, on the one hand, the
hardness of the possible winner problem was established for the
simplified variant of Bucklin's rule only, and on the other hand, we can
show that \textsc{BV-CUSB} is $\np$-complete even for elections with
just two voters.

\begin{theorem}\label{thm:bv-cusb}
  \textsc{BV-CUSB} is $\np$-complete, even for elections with two
  voters.
\end{theorem}
\begin{proofs}
  It is easy to see that \textsc{BV-CUSB} is in $\np$. We show
  $\np$-hardness by a reduction from the following problem (which we will
  refer to as \textsc{Single-Vote Swap Bribery}): Given a vote $v$
  (expressed as a preference order over some candidate set $C$), a
  swap-bribery price function $\pi$ for $v$, a designated candidate $p
  \in C$, and two nonnegative integers $\ell$ and $k$, is there a
  sequence of swaps of adjacent candidates, of total cost at most $k$,
  that ensures that $p$ is ranked among the top $\ell$ positions in $v$.
  (Elkind et al.~\cite{elk-fal-sli:c:swap-bribery} studied this
  problem as a variant of the swap bribery problem for $k$-approval
  elections, where $k$ is part of the input and the election consists
  of a single vote; they established $\np$-completeness of the problem
  in their Theorem~6.)

  Let $I = (C,v,\pi,p,\ell,k)$ be an instance of \textsc{Single-Vote
    Swap Bribery}. We form a Bucklin election $E = (A,V)$ as follows.
  Let $C'$ be a collection of some $\|C\|-1$ dummy candidates. We set
  $A = C \cup C' \cup \{d\}$.  We partition $C'$ into two sets, $C'_1$
  and $C'_2$, such that $\|C'_1\| = \ell$ and $\|C'_2\| =
  \|C'\|-\ell$. (We pick any easily computable partition.)  We let $V$
  be a collection of two voters, $v_1$ and $v_2$, with price functions
  $\pi_1$ and $\pi_2$:
  \begin{enumerate}
  \item $v_1$ has preference order $d > v > C'$ (i.e., $v_1$ ranks $d$
    on the top position, then all the candidates from $C$ in the same
    order as $v$, and then all the candidates from $C'$, in some
    arbitrary-but-easy-to-compute order). For each two candidates $x,y
    \in A$, if both $x$ and $y$ are in $C$ then we set $\pi_1(x,y) =
    \pi(x,y)$, and otherwise we set $\pi_1(x,y) = k+1$.

  \item $v_2$ has preference order $p > C'_1 > d > C'_2 > C - \{p\}$
    (that is, $v_2$ ranks $p$ first, then $\ell$ candidates from
    $C'_1$ followed by $d$, followed by the remaining candidates from
    $C'$, and, then, followed by the candidates from $C-\{p\}$). For each
    two candidates $x,y \in A$, we set $\pi_2(x,y) = k+1$.
  \end{enumerate}
  Note that in our election $\maj{V} = 2$ and, if $p$ is not among the
  top $\ell$ positions within $v$, $d$ is a winner with Bucklin score
  $\ell+2$ (we cannot say that $d$ is the unique winner because we do
  not know on what position $p$ is ranked in $v$).  We claim that $p$
  can become a unique Bucklin winner of election $E$ through a swap
  bribery of cost at most $k$ if and only if $I$ is a yes-instance of
  \textsc{Single-Vote Swap Bribery}.

  Assume that $I$ is a yes-instance of \textsc{Single-Vote Swap
    Bribery}.  This means that there is a sequence of swaps within $v$
  after which $p$ is ranked among the top $\ell$ positions in
  $v$. Applying the same swaps to $v_1$ would cost the same and would
  put $p$ among top $\ell+1$ positions in $v_1$, making $p$ the unique
  Bucklin winner.

  On the other hand, assume that there is a cost-at-most-$k$ sequence
  of swaps within $V$ that make $p$ a unique Bucklin winner. Since any
  swap that is not in the $v$ part of $v_1$ costs $k+1$, we have that
  $d$'s Bucklin score is still $\ell+2$, and, thus, after the swaps,
  $p$'s Bucklin score is in $\{2, \ldots, \ell+1\}$. Executing the
  same swaps within $v$ shows that $I$ is a yes-instance of
  \textsc{Single-Vote Swap Bribery}.~\end{proofs}

To establish that \textsc{BV-DUSB} also is $\np$-complete for the case
of two voters, it suffices to use the same construction as above, with
the exception that now (a) $d$ is the designated candidate whose
victory we want to preclude, and (b) $v_2$ ranks $d$ on position
$\ell+1$ (and not $\ell+2$). Analogous results for the fallback rule
follow immediately.

\begin{corollary}\label{thm:bv-fv-usb}
  \textsc{BV-DUSB}, \textsc{FV-CUSB}, and \textsc{FV-DUSB} are
  $\np$-complete even for the cases of two voters.
\end{corollary}

\subsection{Results for Extension Bribery}

Let us now move on to the study of extension bribery. The following
observation will simplify our discussion.

\begin{observation}\label{obs:eb}
  In (constructive) extension bribery problems for the fallback rule it is
  never profitable to extend any vote in any other way than by asking
  the voter to include the designated candidate on the last unranked
  position.
\end{observation}
Thus we will often specify the extension bribery price functions by
simply giving the cost of extending the vote by just one candidate (we
will refer to this number as \emph{extension cost} of the vote).

Not surprisingly, the weighted variants of extension bribery are
$\np$-complete.

\begin{theorem}
\label{thm:fv-w-p-micro}
Both {\sc FV-CWEB} and {\sc FV-DWEB} are $\np$-complete.
%
\end{theorem}

\begin{proofs}
Obviously,
{\sc FV-CWEB} is in $\np$.  To show $\np$-hardness, we use a
reduction from {\sc Partition}.  Let
$(\{1,\ldots,k\},(a_1,\ldots,a_k))$ be an instance of {\sc Partition}.
We define a fallback election $(C,V)$ with the candidate set $C =
\{b,c,p\}$, the designated candidate~$p$, and $V$ consisting of the
following $k+2$ voters:
\begin{enumerate}
\item There is one voter $v_0$ with the ballot
$
p\,\mid\,\{b,c\},
$
with weight $K+1$ and extension cost $K+1$.

\item For each $i, 1\leq i \leq k$, there is a voter $v_i$ who casts the
  ballot
$
c\,\mid \,\{b,p\},
$
has weight $w_i = a_i$, and has extension cost $a_i$.

\item There is one voter $v_{k+1}$ who casts the ballot
$
b \, \mid \, \{c,p\}
$
with weight $2K$ and extension cost $K+1$.
\end{enumerate}

The total sum of the voter's weights in this election is $5K+1$, so
$\weightedmaj(V) > 2K$.  The weighted scores of the candidates in
$(C,V)$ are shown in Table~\ref{tab:fv-w-p-micro(C,V)-original}.
Both $c$ and $b$ are fallback winners in $(C,V)$ and they win
by approval, thus $p$ is not a (unique) fallback winner in $(C,V)$. 

\begin{table}[h!t]
\captionsetup{font=small}
\caption{Scores in the election constructed in the proof of
 Theorem~\ref{thm:fv-w-p-micro}}
\label{tab:fv-w-p-micro(C,V)}
\centering
\subfigure[Total scores in $(C,V)$]{
\label{tab:fv-w-p-micro(C,V)-original}
\begin{tabular}{@{} l c c c @{}}
\toprule
		&	 $b$ 	& $c$	& $p$	\\
\cmidrule(r){1-1} \cmidrule(rl){2-2} \cmidrule(rl){3-3}  \cmidrule(l){4-4}
$\weightedscore^{1}$ & $2K$	& $2K$ 	& $K+1$\\
\bottomrule	 
\end{tabular}
}
\qquad
\subfigure[Total scores in $(C,V')$]{
\label{tab:fv-w-p-micro(C,V)-modified}
\begin{tabular}{@{} l c c c @{}}
\toprule
			& $b$ 	& $c$	& $p$	\\
\cmidrule(r){1-1} \cmidrule(rl){2-2} \cmidrule(rl){3-3} \cmidrule(l){4-4} 
 $\weightedscore^{1}$ & $2K$	& $2K$ 	& $K+1$	\\
 $\weightedscore^{2}$ & $2K$ 	& $2K$ 	& $2K+1$	\\
\bottomrule	 
\end{tabular}
}
\end{table}

%

We claim that there is a set $A'\seq A= \{1,\ldots , k\}$ such that
$\sum_{i\in A'}a_i = \sum_{i\not\in A'}a_i = K$ if and only if $p$ can
be made the unique fallback winner by extension-bribing some of the voters
without exceeding the budget~$K$.

From left to right: Suppose that there is a set $A' \seq A$ such that
$\sum_{i\in A'}a_i = \sum_{i\not\in A'}a_i = K$.  Change the votes of
those voters $v_i$ with $i \in A'$ from $c\,\mid \,\{b,p\}$ to
$c > p\,\mid \,\{b\}$.  Each of these changes costs $a_i$, so the total
cost is~$K$.  The candidates' scores in the resulting election
$(C,V')$ are shown in Table~\ref{tab:fv-w-p-micro(C,V)-modified}.
We see that $p$ is the unique fallback winner in the bribed election.

%

From right to left: Suppose that $p$ is the unique fallback winner in the
election $(C,V')$, where $V'$ is the changed voter set and the
corresponding changes cost at most~$K$.
Hence, the only changes that can be made (and that follow
Observation~\ref{obs:eb}) are adding the candidate $p$ to the approval
strategies of some of the voters $v_1,\ldots,v_k$.  The scores of the
candidates $b$ and $c$ cannot be decreased, so $p$ has to gain $K$
points to have strictly more points than $b$ and~$c$.  Thus, there
exists a set $A' \seq A$ such that $\sum_{i\in A'}a_i = \sum_{i\not\in
  A'}a_i = K$ and $p$ has to be added to the approval strategies of
those voters $v_i$ with $i \in A'$.

The destructive case can be proven by changing the role of candidates
$p$ and $c$ and changing the weights of both $v_0$ and $v_{k+1}$
to~$K$.~\end{proofs}

On the other hand, the unweighted variant of the problem is in $\p$.
This is a nice complement to the hardness results of Schlotter et
al.~\cite{sch-fal-elk:c:campaign-management-under-approval-driven-voting}
regarding support bribery. The main difference regarding support
bribery and extension bribery is that under the former we assume the
voters to rank all the candidates but declare as approved only some of
their top candidates, whereas in the latter (and, in general, in our
model) we assume the voters to rank only the approved candidates and completely
disregard the disapproved candidates.

\begin{algorithm}  
\small
\DontPrintSemicolon
\SetKwInOut{Input}{input}\SetKwInOut{Output}{output}
\Input{$C$ set of candidates\\ 
~$V$ list of voters\\
~$\Delta = (\delta_1, \ldots, \delta_n)$ list of extension bribery price functions\\
~$k$ budget\\
~$p$ designated candidate}
\Output{``YES'' if $(C,V,\Delta,k,p)\in\text{fallback-\textsc{CUEB}}$ \\
~``NO'' if $(C,V,\Delta,k,p)\notin\text{fallback-\textsc{CUEB}}$}
\BlankLine
\ForEach{$s \in \{1, \ldots, \|C\|\}$}
{
  let $(v'_1, \ldots, v'_r)$ be a sublist of $V$ containing votes
  that approve at most $s-1$ candidates and do not approve $c$, sorted
  by extension costs in ascending order;

  \ForEach{$t \in \{0, \ldots, r\}$}
  {
    \uIf{changing $v'_1, \ldots v'_t$ to approve $p$ makes $p$ the unique winner}
       {\uIf{the sum of extension costs of $v'_1, \ldots, v'_t$ is less thank $k$} {return ``YES'';}}
  }
}
return ``NO'';

\caption{Algorithm for Bucklin-\textsc{CUEB}}
\label{alg:fv-cueb}
\end{algorithm}

\begin{theorem}
\label{thm:fv-cueb-dueb}
  \textsc{FV-CUEB} and \textsc{FV-DUEB} are in $\p$.
\end{theorem}
\begin{proofs}
  Let us consider \textsc{FV-CUEB} first.  We claim that
  Algorithm~\ref{alg:fv-cueb} solves the problem in polynomial time.
  The algorithm considers each round $s$ in which $p$ could possibly
  become the unique winner and tries the cheapest bribery that might
  achieve it. The algorithm clearly runs in polynomial time and its
  correctness follows by Observation~\ref{obs:eb}.

  It is clear how to adapt Algorithm~\ref{alg:fv-cueb} to the case of
  nonunique winners. Then, to solve the destructive variant of the
  problem it suffices to check if any candidate other than $p$ can be
  made a nonunique winner within the budget.
\end{proofs}

\bibliographystyle{alpha}
\bibliography{bucklin}

\newcommand{\etalchar}[1]{$^{#1}$}
\begin{thebibliography}{BEH{\etalchar{+}}10}

\bibitem[BCE13]{bra-con-end:b:comsoc}
F.~Brandt, V.~Conitzer, and U.~Endriss.
\newblock Computational social choice.
\newblock In G.~Weiß, editor, {\em Multiagent Systems}, pages 213--283. MIT
  Press, second edition, 2013.

\bibitem[BEH{\etalchar{+}}10]{bau-erd-hem-hem-rot:b:computational-aspects-of-approval-voting}
D.~Baumeister, G.~Erd\'{e}lyi, E.~Hemaspaandra, L.~Hemaspaandra, and J.~Rothe.
\newblock Computational aspects of approval voting.
\newblock In J.~Laslier and R.~Sanver, editors, {\em Handbook on Approval
  Voting}, chapter~10, pages 199--251. Springer, 2010.

\bibitem[BF78]{bra-fis:j:approval-voting}
S.~Brams and P.~Fishburn.
\newblock Approval voting.
\newblock {\em American Political Science Review}, 72(3):831--847, 1978.

\bibitem[BF83]{bra-fis:b:approval-voting}
S.~Brams and P.~Fishburn.
\newblock {\em Approval Voting}.
\newblock Birkh{\"{a}}user, Boston, 1983.

\bibitem[BFLR12]{bau-fal-lan-rot:c:campaigns-for-lazy-voters}
D.~Baumeister, P.~Faliszewski, J.~Lang, and J.~Rothe.
\newblock Campaigns for lazy voters: {Truncated} ballots.
\newblock In {\em Proceedings of the 11th International Joint Conference on
  Autonomous Agents and Multiagent Systems}, pages 577--584. IFAAMAS, June
  2012.

\bibitem[BO91]{bar-orl:j:polsci:strategic-voting}
J.~{Bartholdi III} and J.~Orlin.
\newblock Single transferable vote resists strategic voting.
\newblock {\em Social Choice and Welfare}, 8(4):341--354, 1991.

\bibitem[BS06]{bra-san:j:critical-strategies-under-approval}
S.~Brams and R.~Sanver.
\newblock Critical strategies under approval voting: {W}ho gets ruled in and
  ruled out.
\newblock {\em Electoral Studies}, 25(2):287--305, 2006.

\bibitem[BS09]{bra-san:j:preference-approval-voting}
S.~Brams and R.~Sanver.
\newblock Voting systems that combine approval and preference.
\newblock In S.~Brams, W.~Gehrlein, and F.~Roberts, editors, {\em The
  Mathematics of Preference, Choice, and Order: {Essays} in Honor of {Peter}
  {C.} {Fishburn}}, pages 215--237. Springer, 2009.

\bibitem[BTT89]{bar-tov-tri:j:manipulating}
J.~{Bartholdi III}, C.~Tovey, and M.~Trick.
\newblock The computational difficulty of manipulating an election.
\newblock {\em Social Choice and Welfare}, 6(3):227--241, 1989.

\bibitem[BTT92]{bar-tov-tri:j:control}
J.~{Bartholdi III}, C.~Tovey, and M.~Trick.
\newblock How hard is it to control an election?
\newblock {\em Mathematical and Computer Modelling}, 16(8/9):27--40, 1992.

\bibitem[CSL07]{con-san-lan:j:when-hard-to-manipulate}
V.~Conitzer, T.~Sandholm, and J.~Lang.
\newblock When are elections with few candidates hard to manipulate?
\newblock {\em Journal of the ACM}, 54(3):Article~14, 2007.

\bibitem[DS12]{dor-sch:j:parameterized-swap-bribery}
B.~Dorn and I.~Schlotter.
\newblock Multivariate complexity analysis of swap bribery.
\newblock {\em Algorithmica}, 64(1):126--151, 2012.

\bibitem[EF10a]{elk-fal:c:shift-bribery}
E.~Elkind and P.~Faliszewski.
\newblock Approximation algorithms for campaign management.
\newblock In {\em Proceedings of the 6th International Workshop On Internet And
  Network Economics}, pages 473--482. Springer-Verlag {\it Lecture Notes in
  Computer Science \#6484}, December 2010.

\bibitem[EF10b]{erd-fel:c:fallback-voting}
G.~Erd\'{e}lyi and M.~Fellows.
\newblock Parameterized control complexity in {Bucklin} voting and in fallback
  voting.
\newblock In V.~Conitzer and J.~Rothe, editors, {\em Proceedings of the 3rd
  International Workshop on Computational Social Choice}, pages 163--174.
  Universit{\"{a}}t D{\"{u}}sseldorf, September 2010.

\bibitem[EFRS12]{erd-fel-pir-rot:t-with-AAMAS-12-pointer:control-in-bucklin-and-fallback-voting}
G.~Erd\'{e}lyi, M.~Fellows, J.~Rothe, and L.~Schend.
\newblock Control complexity in {Bucklin} and fallback voting.
\newblock Technical Report arXiv:{\allowbreak}1103.2230 [cs.CC], Computing
  Research Repository, \mbox{arXiv.org/corr/}, March 2012.
\newblock March, 2011. Revised August, 2012. Extends the AAMAS-2011 paper
  \cite{erd-pir-rot:c:voter-partition-in-bucklin-and-fallback-voting}.

\bibitem[EFS09]{elk-fal-sli:c:swap-bribery}
E.~Elkind, P.~Faliszewski, and A.~Slinko.
\newblock Swap bribery.
\newblock In {\em Proceedings of the 2nd International Symposium on Algorithmic
  Game Theory}, pages 299--310. Springer-Verlag {\it Lecture Notes in Computer
  Science~\#5814}, October 2009.

\bibitem[ENR09]{erd-now-rot:j:sp-av}
G.~Erd\'{e}lyi, M.~Nowak, and J.~Rothe.
\newblock Sincere-strategy preference-based approval voting fully resists
  constructive control and broadly resists destructive control.
\newblock {\em Mathematical Logic Quarterly}, 55(4):425--443, 2009.

\bibitem[EPR11]{erd-pir-rot:c:voter-partition-in-bucklin-and-fallback-voting}
G.~Erd\'{e}lyi, L.~Piras, and J.~Rothe.
\newblock The complexity of voter partition in {Bucklin} and fallback voting:
  {Solving} three open problems.
\newblock In {\em Proceedings of the 10th International Joint Conference on
  Autonomous Agents and Multiagent Systems}, pages 837--844. IFAAMAS, May 2011.

\bibitem[ER10]{erd-rot:c:fallback-voting}
G.~Erd\'{e}lyi and J.~Rothe.
\newblock Control complexity in fallback voting.
\newblock In {\em Proceedings of Computing: the 16th Australasian Theory
  Symposium}, pages 39--48. Australian Computer Society {\it Conferences in
  Research and Practice in Information Technology Series}, vol.~32, no.~8,
  January 2010.

\bibitem[FHH09]{fal-hem-hem:j:bribery}
P.~Faliszewski, E.~Hemaspaandra, and L.~Hemaspaandra.
\newblock How hard is bribery in elections?
\newblock {\em Journal of Artificial Intelligence Research}, 35:485--532, 2009.

\bibitem[FHHR09]{fal-hem-hem-rot:j:llull-copeland-full-techreport}
P.~Faliszewski, E.~Hemaspaandra, L.~Hemaspaandra, and J.~Rothe.
\newblock Llull and {Copeland} voting computationally resist bribery and
  constructive control.
\newblock {\em Journal of Artificial Intelligence Research}, 35:275--341, 2009.

\bibitem[FP10]{fal-pro:j:manipulation}
P.~Faliszewski and A.~Procaccia.
\newblock {AI}'s war on manipulation: {A}re we winning?
\newblock {\em AI Magazine}, 31(4):53--64, 2010.

\bibitem[GJ79]{gar-joh:b:int}
M.~Garey and D.~Johnson.
\newblock {\em Computers and Intractability: A Guide to the Theory of
  NP-Completeness}.
\newblock {W. H. Freeman and Company}, 1979.

\bibitem[HHR07]{hem-hem-rot:j:destructive-control}
E.~Hemaspaandra, L.~Hemaspaandra, and J.~Rothe.
\newblock Anyone but him: {T}he complexity of precluding an alternative.
\newblock {\em Artificial Intelligence}, 171(5--6):255--285, 2007.

\bibitem[KL05]{kon-lan:c:incomplete-prefs}
K.~Konczak and J.~Lang.
\newblock Voting procedures with incomplete preferences.
\newblock In {\em Proceedings of the Multidisciplinary IJCAI-05 Workshop on
  Advances in Preference Handling}, pages 124--129, July/August 2005.

\bibitem[Men11]{men:t-v2:normalized-range-voting-broadly-resists-control}
C.~Menton.
\newblock Normalized range voting broadly resists control.
\newblock Technical Report arXiv:1005.5698v2~[cs.GT], Computing Research
  Repository, \mbox{arXiv.org/corr/}, April 2011.
\newblock Revised June, 2012. To appear in {\it Theory of Computing Systems}.

\bibitem[Pap95]{pap:b:complexity}
C.~Papadimitriou.
\newblock {\em Computational Complexity}.
\newblock Addison-Wesley, second edition, 1995.

\bibitem[Rot05]{rot:b:cryptocomplexity}
J.~Rothe.
\newblock {\em Complexity Theory and Cryptology. An Introduction to
  Cryptocomplexity}.
\newblock EATCS Texts in Theoretical Computer Science. Springer-Verlag, 2005.

\bibitem[RS]{rot-sch:j-toappear:survey-typical-case-challenges-manipulation-control}
J.~Rothe and L.~Schend.
\newblock Challenges to complexity shields that are supposed to protect
  elections against manipulation and control: {A} survey.
\newblock {\em Annals of Mathematics and Artificial Intelligence}.
\newblock To appear.

\bibitem[RS12]{rot-sch:c:fallback-voting-experiments}
J.~Rothe and L.~Schend.
\newblock Control complexity in {Bucklin}, fallback, and plurality voting: {An}
  experimental approach.
\newblock In {\em Proceedings of the {\it 11th International Symposium on
  Experimental Algorithms}}, pages 356--368. Springer-Verlag {\it Lecture Notes
  in Computer Science~\#7276}, June 2012.

\bibitem[SFE11]{sch-fal-elk:c:campaign-management-under-approval-driven-voting}
I.~Schlotter, P.~Faliszewski, and E.~Elkind.
\newblock Campaign management under approval-driven voting rules.
\newblock In {\em Proceedings of the 25th AAAI Conference on Artificial
  Intelligence}, pages 726--731, August 2011.

\bibitem[XC11]{con-xia:j:possible-necessary-winners}
L.~Xia and V.~Conitzer.
\newblock Determining possible and necessary winners given partial orders.
\newblock {\em Journal of Artificial Intelligence Research}, 41:25--67, 2011.

\bibitem[Xia12]{xia:c:margin-of-victory}
L.~Xia.
\newblock Computing the margin of victory for various voting rules.
\newblock In {\em Proceedings of the 13th ACM Conference on Electronic
  Commerce}, pages 982--999. ACM Press, 2012.

\bibitem[XZP{\etalchar{+}}09]{xia-zuc-pro-con-ros:c:unweighted-coalitional-manipulation}
L.~Xia, M.~Zuckerman, A.~Procaccia, V.~Conitzer, and J.~Rosenschein.
\newblock Complexity of unweighted coalitional manipulation under some common
  voting rules.
\newblock In {\em Proceedings of the 21st International Joint Conference on
  Artificial Intelligence}, pages 348--353. IJCAI, July 2009.

\end{thebibliography}

\end{document}